\documentclass[a4paper,12pt]{article}

\usepackage[utf8]{inputenc}
\usepackage{graphicx}
\usepackage{subcaption}
\usepackage{amsmath, amssymb}
\usepackage{booktabs}
\usepackage{geometry}
\usepackage{authblk}  
\usepackage{hyperref} 
\usepackage[numbers]{natbib} 

\title{{Influence of the particle morphology on the spray characteristics in low-pressure cold gas process} \footnotetext{Manuscript submitted on \today.}}

\author[1]{Yannik Sinnwell\thanks{Corresponding author: \href{mailto:yannik.sinnwell@mv.rptu.de}{yannik.sinnwell@mv.rptu.de}}}
\author[1,2]{Anton Maksakov}
\author[2]{Stefan Palis}
\author[1]{Sergiy Antonyuk}

\affil[1]{Institute of Particle Process Engineering, University of Kaiserslautern-Landau (RPTU), Gottlieb-Daimler Straße 44, 67663 Kaiserslautern, Germany}
\affil[2]{Institute of Electrical Engineering, TU Clausthal, Leibnizstraße 28, 38678 Clausthal-Zellerfeld, Germany}

\date{}  

\begin{document}

\maketitle

\begin{abstract}
    This study investigates the influence of particle morphology on spray characteristics in low-pressure cold gas spraying (LPCGS) by analyzing three copper powders with distinct shapes and microstructures. A comprehensive morphology analysis was conducted using both 2D and 3D imaging techniques. Light microscopy combined with image processing quantified particle circularity in 2D projections, while X-ray micro-computed tomography (µCT) enabled precise 3D reconstructions to determine sphericity, surface area, and volume distributions. The results showed significant variations in the particle morphology of the investigated feedstock copper powders, with irregularly shaped particles exhibiting lower circularity and sphericity compared to more spherical feedstocks. These morphological differences had a direct impact on the particle velocity distributions and spatial dispersion within the spray jet, as measured by high-speed particle image velocimetry. Irregular particles experienced stronger acceleration and exhibited a more focused spray dispersion, whereas spherical particles reached lower maximum velocities and showed a wider dispersion in the jet. These findings highlight the critical role of particle morphology in optimization of cold spray processes for advanced coating and additive manufacturing applications.
\end{abstract}

\textbf{Keywords:} Low-pressure Cold Spray; Particle Morphology Analysis; CT Image Analysis; High-Speed Particle Image Velocimetry.

\section{Introduction}
\label{sec1}
The low-pressure Cold Gas Spraying (LPCGS) process is a solid-state coating technique that uses a high-velocity gas stream to accelerate powder particles toward a substrate to deposit them. In contrast to thermal spray methods, LPCGS operates at relatively low temperatures~($T~<~600$°C) and low pressures ~($p~<~15$~bar), allowing particles to adhere without melting~\citep{Assadi.2016, Raoelison.2017, Yin.2021, Papyrin.2007, Champagne.2007}. This preserves the original material properties of the feedstock, making it suitable for a variety of applications, including coating~\citep{Moridi.2014, Aleksieieva.2024}, repair~\citep{Champagne.2015, Aleksieieva.2022} and the additive manufacturing~\citep{Ashokkumar.2022,Oyinbo.2019} of materials, such as metals, polymers and ceramics~\citep{Falco.2025}.\\
In cold spray, a compressed gas (usually air, nitrogen, or helium) is used to accelerate powder particles through a specially designed nozzle, typically a Laval nozzle~\cite{Dykhuizen.1998, Buhl.2018}, over which the gas expands, achieving supersonic velocities. The powder particles are introduced into the gas stream and accelerated to velocities in the range of 200-1200~$m\cdot s^{-1}$, depending on the material and process parameters (pressure $p$ and temperature $T$ of the gas), as well as the nozzle geometry). During impact on the surface, the particles deform plastically and bond due to mechanical interlocking and adiabatic shear instability in contact~\cite{Dykhuizen.1999, Grujicic.2003}. \\
The process is largely governed by impact energy, and successful deposition occurs when particles reach a material-dependent critical impact velocity~$v_{crit}$~\cite{Dykhuizen.1999, Assadi.2003}. It is mainly governed by material properties, such as the ultimate tensile strength, the material density, the temperature of the contact partners and the specific heat capacities of their materials~\cite{Assadi.2003, Assadi.2011, Schmidt.2006}. Furthermore, the particle shape influences the critical velocity, as this determines the stresses and deformations resulting from the impact~\citep{Palodhi.2021}. \\
The impact velocity of the particles is directly related to the velocity of the particles in the spray jet. The acceleration of the particles, and thus their velocity in the spray, is primarily determined by the drag force in Eq.(\ref{eq_drag}), which is particularly dependent on the size and the shape of the particles (cf.~\citep{Haider.1989, Holzer.2008}). 
\begin{equation}
\label{eq_drag}
    \rho_p\cdot V_p\cdot \frac{dv_p}{dt} = F_W = c_w(Re_p, \Psi)\cdot A_p \cdot \frac{\rho_g}{2} \cdot \left(v_g - v_p \right)^2
\end{equation}
A similar relationship exists between the impact velocity of the particles and the drag force experienced during penetration through the gas stagnation zone above the substrate, which results in a reduction in velocity~\citep{Grujicic.2004, Dykhuizen.1998}. The velocities of the particles in the spray jet and their impact velocity on the substrate can be analysed using numerical simulation. Euler-Lagrange Computational Fluid Dynamics (CFD) simulations are utilised for this purpose as summarized in the review of Yin et al.~\citep{Yin.2016}. The review also provides a summary of the literature relating to the experimental analysis of spray properties when using different particle morphologies. Wang et al.~\citep{Wang.2022} performed CFD simulations considering the shape of the particles and demonstrated that as the degree of sphericity decreased, the particle stream became wider while the spherical particles became more focused. Varadaraajan and Mohanty~\citep{Varadaraajan.2017} showed the effect of expansion ratio on the gas flow, particle velocities and deposition efficiency by numerical simulation assuming the spherical particle shape and experiments with the stainless steel powder. Lupoi and O'Neill~\citep{Lupoi.2011} conducted CFD~simulations of 20~µm copper particles in the gas flow of a Laval nozzle, showing that the gas and particle velocities increase towards the centre of the spray. Furthermore, it was demonstrated that the particles change their trajectory at the central exit of the nozzle and become more radially distributed in the spray. The deflection of particles was attributed to the high gas turbulence experienced at the nozzle exit, although this analysis was conducted for a single particle size. \\
In particular, the prediction of the particle behavior of irregularly shaped particles proves to be challenging. In our previous works~\cite{Breuninger.2019, Breuninger.2019b, Schmidt.2017, Aleksieieva.2024}, numerical simulations were conducted to evaluate particle impact velocities and temperatures and to define the deposition operating window for various materials. Smaller particles accelerate rapidly, nearly reaching the gas velocity within the nozzle. However, they experience significant deceleration above the surface due to the gas stagnation region. In contrast, larger particles fail to attain high velocities because of their greater inertia. This results in a characteristic velocity maximum within a narrow particle size range, highlighting a pronounced particle separation effect within the spray. The critical impact velocity, computed using the model proposed in~\cite{Assadi.2003}, indicates a narrow deposition window for the selected process parameters and materials. The results emphasize that even minor variations in particle size and shape~\cite{Breuninger.2019, Ning.2007, Song.2017, Gu.2009} can significantly alter impact velocities. Furthermore, studies on the spreading behavior of gas-particle sprays~\cite{Breuninger.2019b} have shown that particle size strongly influences spray width and, consequently, the minimum deposition area impacting the resolution of structure generation. \\
To quantify the influence of particle morphology on spray behavior, different optical in-line measuring systems are used, which allow the analysis of particle velocity, particle position and size~\citep{DelshadKhatibi.2017}. The most common measurement methods are based on interaction with laser light. These include scattered light-based methods and backlight illumination methods. Scattered light methods make it possible to analyze the particle speed and size in a range of 100 to 1500~$m\cdot s^{-1}$ and a particle size range of 10 to 300~µm~\citep{Fukanuma.2006, Neo.2022}. Compared to methods that work with backlight illumination, the spatial resolution is very low due to the punctual measurement method and no particle positions can be determined. In comparison, the backlight illumination method enables a high spatial resolution and can therefore determine not only the particle speed and size but also their position in the spray jet. Furthermore, this method employs a wider measuring range for both particle velocity and size (2 to 2000~$m\cdot s^{-1}$ and 5~µm to 2~mm)~\citep{Larjo.2004, Koivuluoto.2018, Ozdemir.2020}. In this work, a High Speed Particle Image Velocimetry (H-PIV) system with a backlight illumination imaging method (shadow imaging) is used. The particles are detected through the local extinction of a uniform laser light field, and telecentric imaging is used to ensure a good depth of field for accurate particle tracking~\cite{Koivuluoto.2018}. In order to further develop the cold gas process into an autonomous and precise manufacturing process, the relationship between the particle morphology and the spraying behavior must therefore be investigated. Model-based optimization of the cold gas process parameters has already been carried out for this purpose~\citep{WilhelmT.2025}. In the future, these models should take into account the morphological influence of the particle systems. \\ 
A number of authors have already dealt with the influence of particle size and morphology on the properties of sprayed coatings~\citep{Winnicki.2015, Shockley.2015, Munagala.2018, Goral.2019}. It has been demonstrated that both the particle shape and size have an effect on the deposition efficiency and the adhesion of the coating layer to the substrate. For example, irregularly shaped particles have been shown to exhibit a higher deposition efficiency (DE) than spherical particles, but perform worse in terms of adhesion to the substrate. In addition, layers with spherical particles result in a lower coating porosity than layers with irregularly shaped particles~\citep{Wong.2013, Palodhi.2021}. \\ 
A review of the extant literature indicates that the velocity and trajectory angles of particles are influenced by a range of particle properties and their interactions with the gas phase. To specifically investigate the role of particle morphology, this study employs a feedstock material with identical chemical composition, differing solely in particle morphology. The primary objective is to quantify the impact of particle morphology on the velocity distribution and spatial particle distribution within the spray jet. This is achieved using high-speed particle image velocimetry~(H-PIV), with process gas temperature and the distance from the nozzle exit as key variables. Additionally, the parameter study is complemented by a detailed 2D and 3D image-based morphological analysis of the copper particle systems examined~\citep{Pashminehazar.2018, Wang.2025}.


\section{Materials and Methods}
\label{sec2}
\subsection{Morphological analysis of the feedstock powders by means of 2D and 3D image analysis}
\label{subsec2_1}

\begin{figure}  [htbp]
\centering
\begin{subfigure}{0.3\textwidth}
    \includegraphics[width=\textwidth]{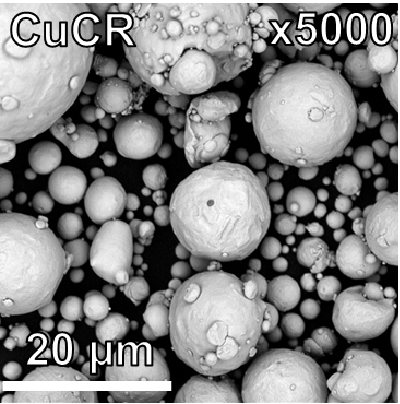}
    \caption{CuCR}
    \label{REM_CUCR}
\end{subfigure}
\hfill
\begin{subfigure}{0.3\textwidth}
    \includegraphics[width=\textwidth]{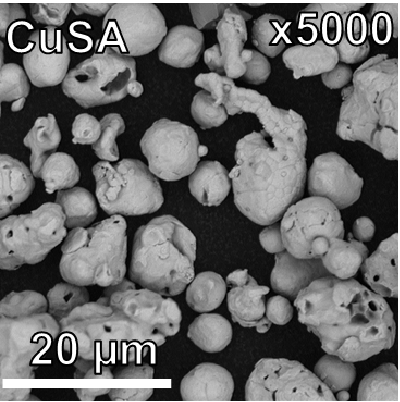}
    \caption{CuSA}
    \label{REM_CUSA}
\end{subfigure}
\hfill
\begin{subfigure}{0.3\textwidth}
    \includegraphics[width=\textwidth]{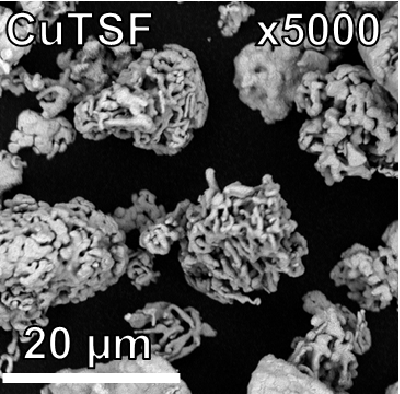}
    \caption{CuTSF}
    \label{REM_CUTSF}
\end{subfigure}
\begin{subfigure}{0.45\textwidth}
    \includegraphics[width=\textwidth]{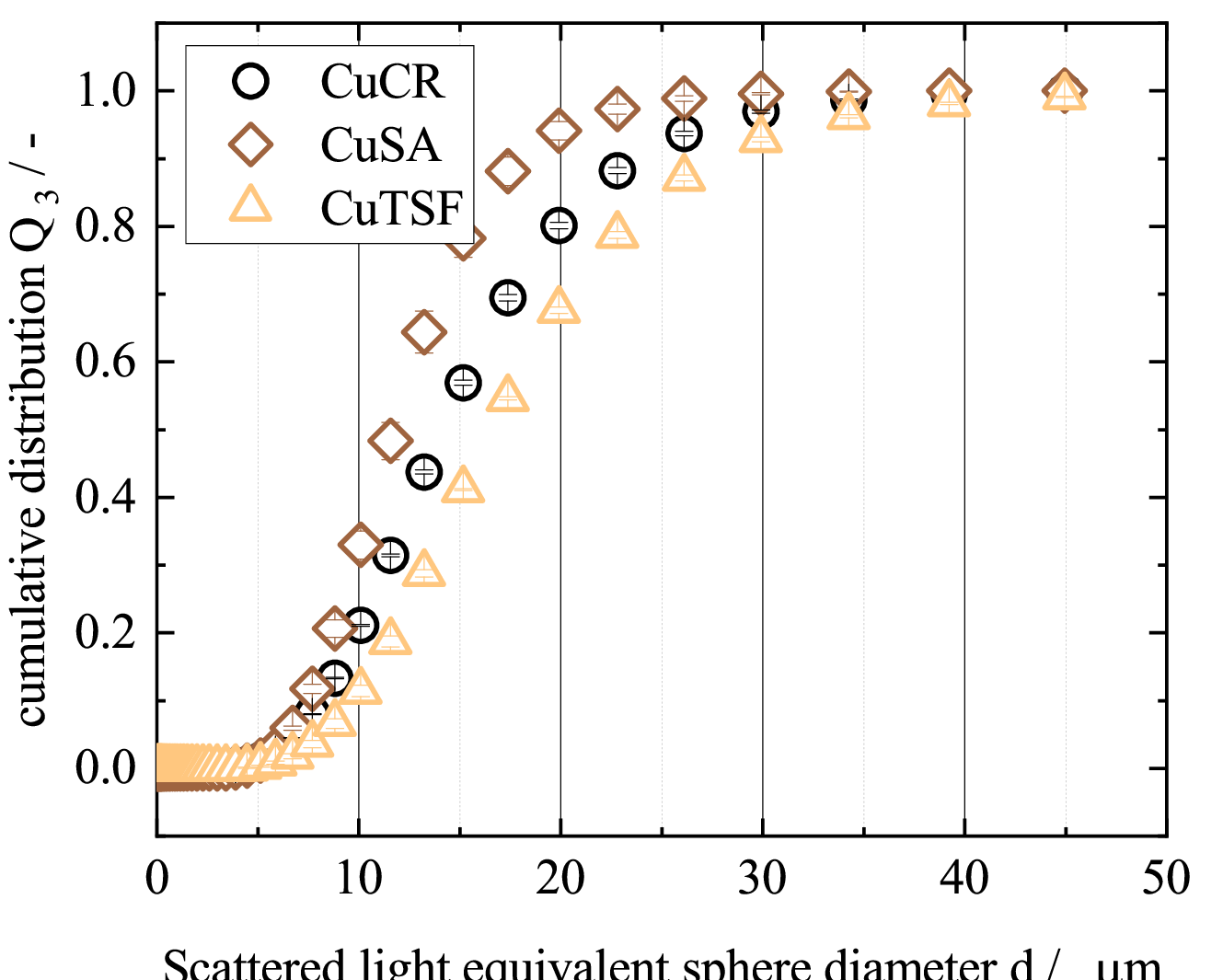}
    \caption{Cumulative Particle Size Distribution}
    \label{PSD}
\end{subfigure}
\caption{Scanning electron microscope images of the copper feedstock powders (a)-(c) and their volume-based cumulative particle size distribution (d).}
\label{subfigure_REM}
\end{figure}

\begin{figure} [htbp]
    \centering
    \includegraphics[width=1.0\linewidth]{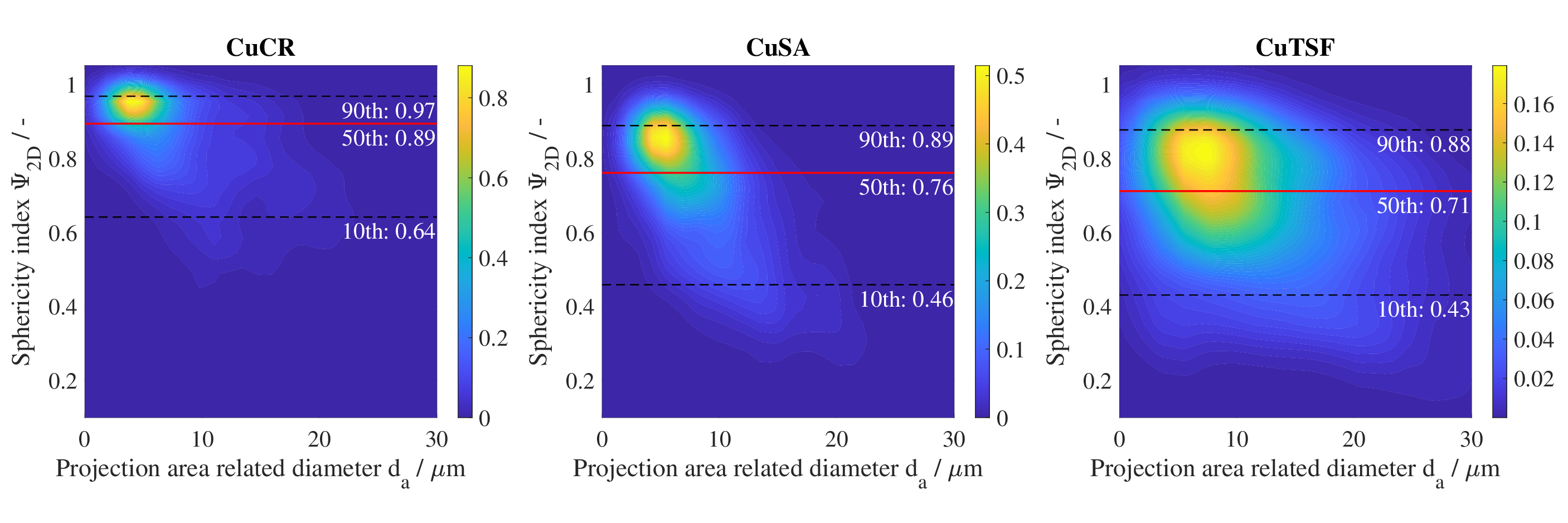}
    \caption{Number-based probability density maps of the circularity index $\Psi_{2D}$ of the analyzed feedstock powders. The horizontal lines indicate the 10th, 50th and 90th percentile of the circularity distribution. The color scale at the right of each graph represents the range of kernel densities.}
    \label{fig_2_1a}
\end{figure}

\begin{figure} [htbp]
    \centering
    \includegraphics[width=0.7\linewidth]{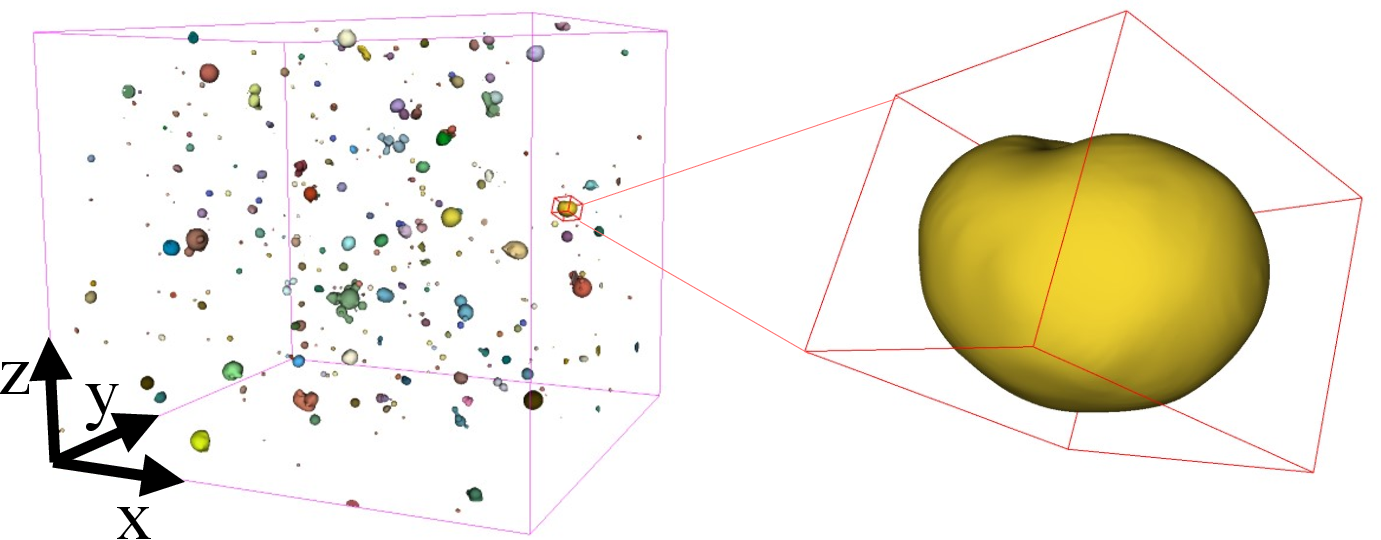}
    \caption{Exemplary CT image based 3D reconstruction of the investigated feedstock powder CuCR.}
    \label{fig_2_2}
\end{figure}

In this study the particle velocities and trajectories of three different copper powders: I) Carl Roth Copper Powder~(CuCR), $X_{Cu}\geq$~99.7~\%, 1~$<d_{p}<$~63~µm; II) Sigma Aldrich Copper Powder~(CuSA), $X_{Cu}\geq$~98.0~\%, 1~$<d_{p}<$~25~µm and III) Fisher Scientific Copper Flake~(CuTSF), $X_{Cu}\geq$~99.0~\%, 1~$<d_{p}<$~45~µm, were investigated. The CuCR and CuTSF powders were sieved with a cutoff mesh size of 25~µm and the undersize grain was used for further investigations to equalize the particle size range of all feedstock powders in order to highlight the influence of the particle morphology in the spray experiments. The particle systems were selected for their similar material composition and particle size range, while exhibiting variation in shape and overall morphology, as evidenced by the SEM images presented in Figure~\ref{subfigure_REM}. All powders have a unimodal particle size distribution. The particle size distribution (PSD) is obtained by means of laser diffraction with a Retsch Horiba particle size analyzer~\citep{Bottlinger.1989, BLOTT.2006, W.Hintz.2011}. The particle size distributions are presented in Figure~\ref{PSD}, and the characteristic values of the particles are given in Table~\ref{tab_1}. The solid densities of the powders shown in Table~\ref{tab_1} were quantified using helium pycnometry (Micro Ultrapyc 1200e, Quantachrome). \\
The CuCR particles have a predominantly spherical structure with some smaller satellite particles adhering to their surface. Like the CuTSF material, the CuSA powder shows more irregular particle shapes with open-pored surface defects, which are more pronounced in the CuTSF material and resemble a sponge-like structure. The particle size distributions demonstrate that the studied materials exhibit comparable particle sizes, with a slight decrease in particle size from CuTSF to CuCR to CuSA particles. Looking at the densities of the powder material, the CuCR and CuTSF powders show a high agreement with bulk copper, whereas the CuSA material shows a slight deviation in the form of a lower density. In view of the surface defects visible in the SEM images, this can be attributed to closed pores in the material~(cf.~Fig.~\ref{REM_CUSA}).
\begin{table}[htbp]
\caption{Size range and material density of the feedstock copper powders.}
\label{tab_1}
\centering
\begin{tabular}{lllll}
\hline
      & \multicolumn{1}{c}{$D_{10,3}$} & \multicolumn{1}{c}{$D_{50,3}$} & \multicolumn{1}{c}{$D_{90,3}$} & \multicolumn{1}{c}{Solid density} \\
      & \multicolumn{3}{c}{µm}                                                      & \multicolumn{1}{c}{$kg \cdot m^{-3}$}    \\ \hline
CuCR  &   9.27 $\pm$ \textit{0.02}                      &   16.13 $\pm$ \textit{0.06}                     &     27.01 $\pm$ \textit{0.27}                    &         8759.1 $\pm$ \textit{0.007}             \\
CuSA  &   7.46 $\pm$ \textit{0.10}                     &   11.91 $\pm$ \textit{0.27}                     &     18.57 $\pm$ \textit{0.73}                  &         8417.5 $\pm$ \textit{0.005}            \\
CuTSF &   9.49 $\pm$ \textit{0.30}                     &   16.61 $\pm$ \textit{0.45}                     &     28.22 $\pm$ \textit{0.36}                  &         8829.2 $\pm$ \textit{0.031}            \\ \hline
\end{tabular}
\end{table}
To quantify the morphology, the particle shape distribution of the respective powders is analyzed. Different imaging methods are used for this purpose, as they allow the elucidation of particle shape, surface morphology in combination with the size distribution~\cite{Merkus.2008, Radvilaite.2016}. Firstly, the particle shape is quantified using light microscopic images (Leitz~Orthoplan, Germany). The images are then processed in MATLAB\textsuperscript{\textregistered} via segmentation and binarization. The \textit{regionprops} function is used to calculate the circularity $\Psi_{2D}$ based on the projection area $A$ of the particles. The circularity $\Psi_{2D}$ describes the roundness of an object's projection region $A$ by relating the object's region-equivalent circle perimeter to its actual perimeter $u$ \citep{Image.Processing.Toolbox.2024}. Equation (\ref{eq_1}) is used to compute the object's circularity. It includes a correction term $r$ to remove the bias that causes circularity values to be too high for relatively small objects as a result of the underestimation of the perimeter calculation by the connection of pixel centers by straight lines \citep{Vossepoel.1982}.  
\begin{equation}
\label{eq_1}
    \Psi_{2D}=\frac{4\cdot\pi\cdot A}{u^{2}}\cdot\left(1 -\frac{0.5}{r}\right)^{2}\textnormal{with }r=\frac{u}{2\cdot\pi}
\end{equation}
The results of the 2D shape analysis are displayed as kernel density contour plots in Figure~\ref{fig_2_1a}, taking into account the circularity~$\Psi_{2D}$ and the projection area related particle diameter. In order to obtain a representative result, it is ensured that at least 2000 individual particles are analyzed as described. The shape distribution results indicate the following decrease in particle projection area circularity: CuCR~$\Psi_{2D,50}=0.89$, CuSA~$\Psi_{2D,50}=0.76$ and CuTSF~$\Psi_{2D,50}=0.71$. With regard to the width of the shape distribution, the CuSA and especially the CuTSF powder show a significantly wider distribution than the CuCR powder. The latter has a comparatively low density distribution width with a high probability density for high circularities. These findings are in agreement with the represented particle collectives in the SEM images in Figure~\ref{subfigure_REM}. All distributions demonstrate a slight tendency to decrease in form factor for larger particle diameters (see~Fig.~\ref{fig_2_1a}).\\
Since the circularity $\Psi_{2D}$ is calculated on the basis of the projection area of the particles and therefore only provides a two-dimensional representation, a three-dimensional shape descriptor is also determined by means of X-ray micro-computed tomography using a TomoscopeL\textsuperscript{\textregistered} from Werth Messtechnik GmbH. For the measurements, the powders are uniformly dispersed in epoxy resin and, after curing, images are taken with a voxel size of 470~nm, which corresponds to the maximum resolution of the setup. The scans are uniformly performed with an acceleration voltage of 150~kV an 8~µA current with 1000 steps to a full probe rotation at an integration time of 8000~ms. This allows for the reconstruction of individual particles with high resolution to determine the complex 3D structure of the raw powders (see~Fig.~\ref{subfigure_REM}). The subsequent image processing with FijI~\citep{Schindelin.2012} and 3D Slicer~\citep{Fedorov.2012} includes voxel scaling of the images, noise suppression by applying a Gaussian blur filter with a radius of $\sigma$~=~2 in all spatial directions, and transformation of the image data into .nrrd format for further processing with 3D Slicer. The segmentation of the copper particles is performed using Otsu's thresholding method~\citep{Otsu.1979}. In a next step, the resulting segment of copper particles is labeled by applying the \textit{'split segments to islands'} function. The resulting 3D reconstruction of the particle collective for CuCR is shown in Figure~\ref{fig_2_2}. The labeled objects allow quantification of individual particle properties, such as volume, particle surface area, and other metrics. The computation of the object's surface is based on the Marching Cube algorithm, which generates a triangulated surface mesh by identifying the boundaries where voxel intensity values change~\citep{Lorensen.1987}. Due to the discrete nature of the voxel grid, the generated mesh includes jagged edges. To account for this, a Laplacian smoothing algorithm is applied~\citep{OlgaSorkine.2005}. The surface area is then computed by the triangular mesh as the sum of the resulting vertices surfaces and since it marks the boundary of the object, the enclosed volume can be calculated using numerical integration methods. The particle shape is described with the sphericity definition by Wadell~\citep{Wadell.1935}, which relates the surface area of the volume $V$ equivalent sphere with the actual surface $S$ of the particle:
\begin{equation}
\label{eq_3}
    \Psi_{3D}=\frac{\left(36\pi V^2\right)^{\frac{1}{3}}}{S}.
\end{equation}
 Figure~(\ref{fig_2_2}) shows an exemplary particle reconstruction as a closed surface mesh (yellow particle) with its Oriented Bounding Box~(OBB). The individual colors of the particle meshes indicate the different labels necessary for the computation of the labeled particle properties (surface area~$S$ and volume~$V$). 

\begin{figure} [htbp]
    \centering
    \includegraphics[width=1.0\linewidth]{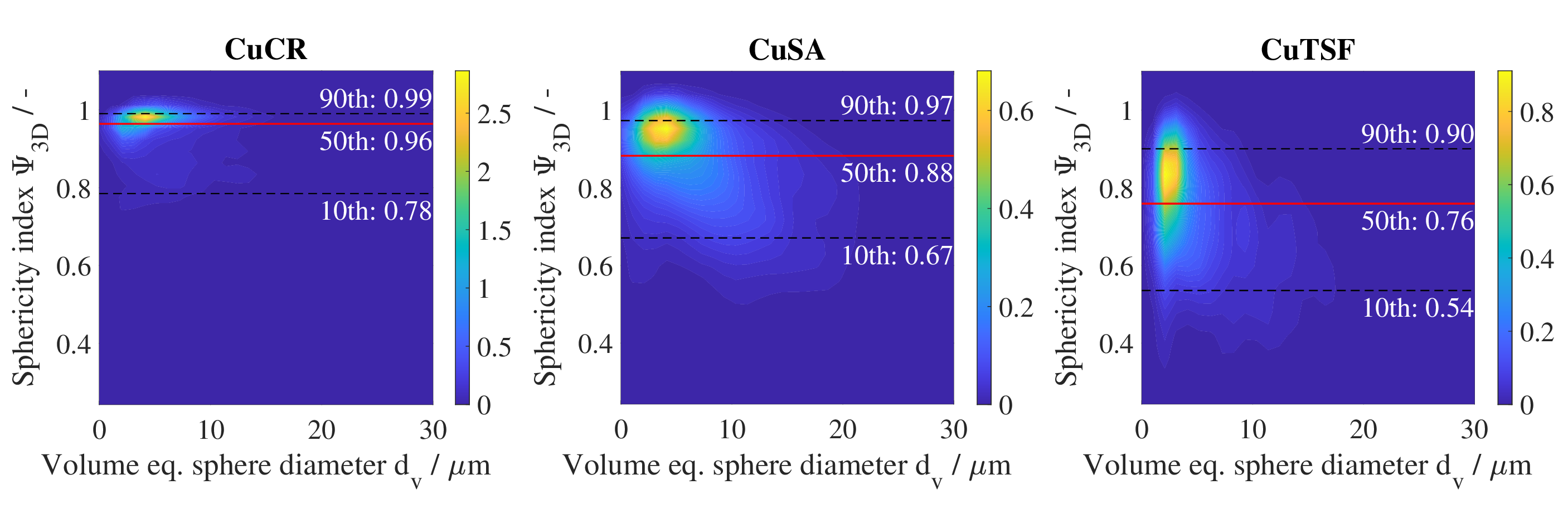}
    \caption{Number-based probability density maps of the Wadell sphericity index $\Psi_{3D}$ of the analyzed feedstock powders. The horizontal lines indicate for the 10th, 50th and 90th percentile of the sphericity distribution.}
    \label{fig_2_3}
\end{figure}
The analysis of the Wadell form factor shows a consistent trend with the results of the 2D form factor analysis. The CuCR powder has a median sphericity index $\Psi_{3D,50}$ of 0.96, followed by CuSA with $\Psi_{3D,50}= 0.88$ and CuTSF with $\Psi_{3D,50}= 0.76$ (cf. Figure~(\ref{fig_2_3})). Compared to the 2D analysis, the measured sphericity show a narrower distribution width, which is due to the capture of the complete particle structure by the 3D data~(cd.~Table~\ref{tab_2}). A detailed comparison of the form factor distribution of the 2D and 3D analysis shows an increase in the form factors for the 3D analysis for the lower edge of the distributions at $\Psi_{10}$ for all particle systems. The same trend can be observed in the opposite direction for the upper edge of the distribution at $\Psi_{90}$. In contrast, the median values of the form factors do not show a trend across particle systems and deviate less strongly between the 2D and 3D analysis. For CuCR the $\Psi_{50}$ values are identical, for CuSA the 3D value increases slightly and for CuTSF it decreases. This is particularly evident when comparing the data for CuTSF. The deviation from the 2D analysis is especially high for the diameter distribution. Compared to the distribution of the projection area equivalent sphere diameter, the distribution of the volume equivalent sphere diameter is significantly narrower and the determined diameters are smaller. All distributions show a slight tendency to decrease in form factor for larger particle diameters (see~Fig.~\ref{fig_2_1a} and Fig.~\ref{fig_2_3}). Furthermore, the density plots show the number distribution of the particle systems. This explains the high densities in the range of low equivalent diameters, which differs from the volume-related particle size distribution in Figure~\ref{PSD}.
\begin{table}[htbp]
\caption{Comparison of the experimentally determined 2D and 3D particle shape descriptor distributions displayed as 90th, 50th, and 10th percentile.}
\label{tab_2}
\centering
\begin{tabular}{lccclcl}
                                                                                & \multicolumn{2}{c}{CuCR}                                                                                                & \multicolumn{2}{c}{CuSA}                                                                                                                    & \multicolumn{2}{c}{CuTSF}                                                                                                                   \\ \cline{2-7} 
                                                                                & 2D                                                         & 3D                                                         & 2D                                                                             & \multicolumn{1}{c}{3D}                                     & 2D                                                                             & \multicolumn{1}{c}{3D}                                     \\ \hline
\begin{tabular}[c]{@{}l@{}}$\Psi_{90}$\\$\Psi_{50}$\\ $\Psi_{10}$\end{tabular} & \begin{tabular}[c]{@{}c@{}}0.97\\ 0.89\\ 0.64\end{tabular} & \begin{tabular}[c]{@{}c@{}}0.99\\ 0.96\\ 0.78\end{tabular} & \multicolumn{1}{l}{\begin{tabular}[c]{@{}l@{}}0.89\\ 0.76\\ 0.46\end{tabular}} & \begin{tabular}[c]{@{}l@{}}0.97\\ 0.88\\ 0.67\end{tabular} & \multicolumn{1}{l}{\begin{tabular}[c]{@{}l@{}}0.88\\ 0.71\\ 0.43\end{tabular}} & \begin{tabular}[c]{@{}l@{}}0.90\\ 0.76\\ 0.54\end{tabular} \\ \hline
\end{tabular}
\end{table}

\begin{figure} [htbp]
    \centering
    \includegraphics[width=1.0\linewidth]{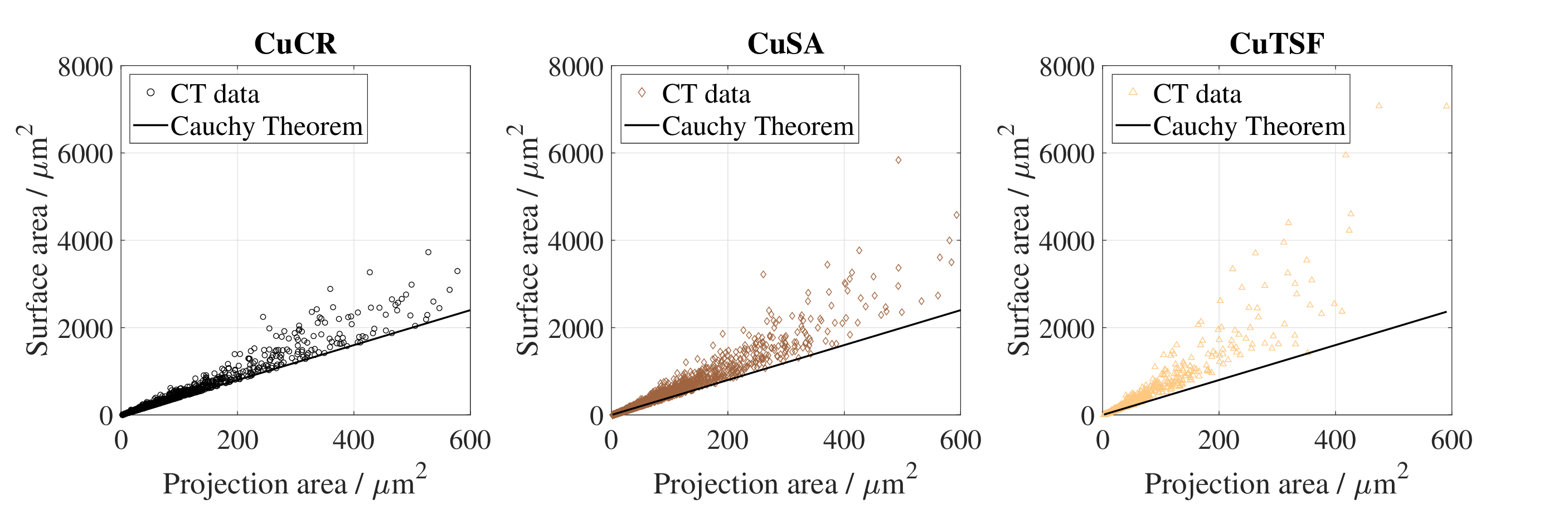}
    \caption{Representation of the particle surface over the projection area calculated using the volume equivalent sphere diameter. For comparison, the dependence of the surface area on the projection area is shown with $S=4\cdot A$ according to Cauchy's theorem.}
    \label{fig_2_1b}
\end{figure}
In addition to form factor analysis, Cauchy's theorem can be employed to examine the investigated particle systems. According to Cauchy's theorem, the average area of the parallel projections of a convex body into the plane is equivalent to one-fourth of its surface area~\citep{AugustinLouisCauchy.1841}. As illustrated in Figure~\ref{fig_2_1b}, the surface area of the particle systems is plotted against the projection area of a volume equivalent sphere. For reference, the course according to Cauchy's theorem is shown. The position of the data points indicates the increasing irregularity of the particle systems from CuCR via CuSA to CuTSF, suggesting that the particles have increasingly concave areas in their surface morphology. This can be observed by examining the SEM images in Figure~\ref{subfigure_REM}. When the morphology of the CuTSF particles is compared with the CuCR and CuSA particles, it can be seen that they have a predominantly porous structure (cf. Fig.~\ref{subfigure_REM}). In addition, a large proportion of the CuTSF particles have a flake shape. This morphology cannot be adequately represented by the projection area evaluation. The discrepancy between the two methods in determining the equivalent diameter can be attributed to the porosity of the CuTSF particles. This can be taken into account when calculating the volume of the particle reconstructions, as the volume-related equivalent diameter shows comparatively low values. Corresponding results can be found in the literature when comparing 2D and 3D measurement methods~\citep{Beemer.2022, Li.2021, Li.2023, Rorato.2019}. 
\begin{figure} [htbp]
    \centering
    \includegraphics[width=1.0\linewidth]{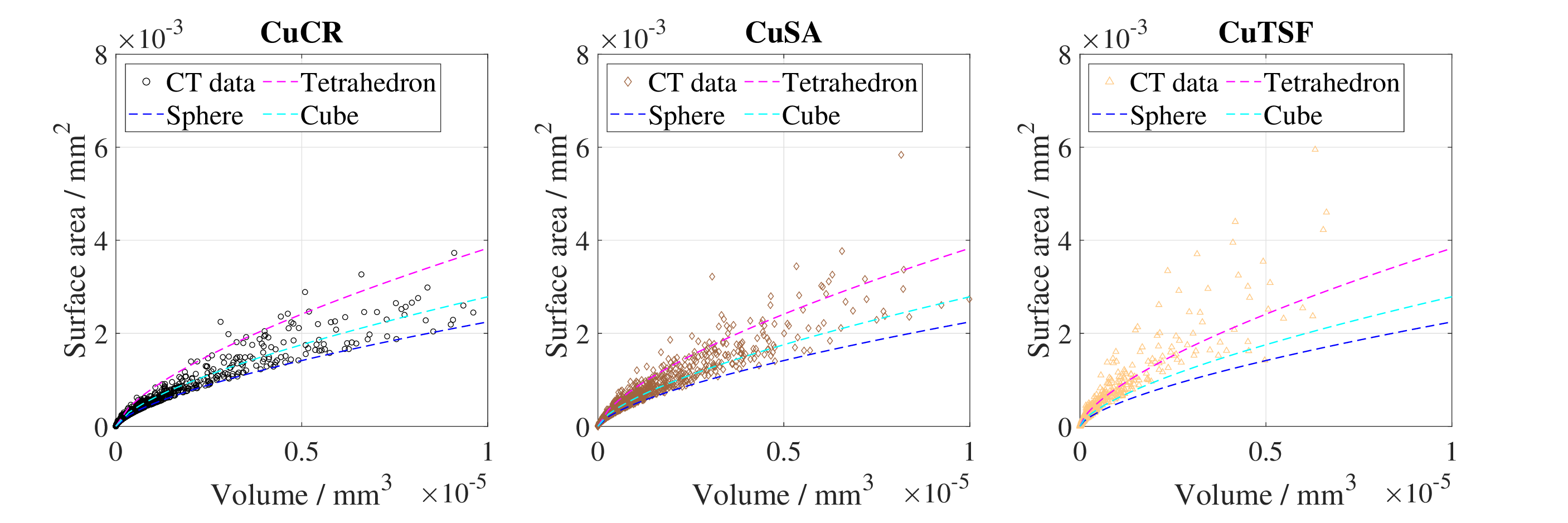}
    \caption{Plots of the reconstructed particle surface against their volume. For comparison, the surface against the volume plots of ideal geometric bodies (tetrahedron, cube and sphere) are shown as dotted lines. The black solid line represents the fit to the experimental data.}
    \label{fig_SAtoV}
\end{figure}
If the measured particle surfaces are plotted over the respective particle volume and the course of the fit is compared with that of ideal symmetrical bodies, it can be determined that a large number of the particles are within the range of these bodies (cf. Figure~(\ref{fig_SAtoV})). The sphere, the cube and the tetrahedron serve as symmetrical reference bodies because they cover a wide range of S/V ratios and sphericity ($\Phi_{sphere}$=1.0; $\Phi_{cube}\approx$ 0.806; $\Phi_{tetrahedron}\approx$ 0.671). In the case of the CuCR and CuSA powders, the majority of the tomographed particles lie between the surface-volume-line plot of the sphere and the tetrahedron. The CuTSF particles, on the other hand, show a stronger deviation from the reference bodies in the direction of higher S/V ratios. This can be attributed to the fact that some of the particles have concave shape features that cannot be reproduced by the convex reference bodies. The shape and size analysis of the particle systems used shows that they are a suitable choice for quantifying the influence of morphology on the behavior in the spray jet.  

\subsection{Experimental setup: Cold gas spraying with high speed particle image velocimetry (H-PIV)}
\label{subsec2_2}

\begin{figure}  [htbp]
    \centering
    \includegraphics[width=1.0\textwidth]{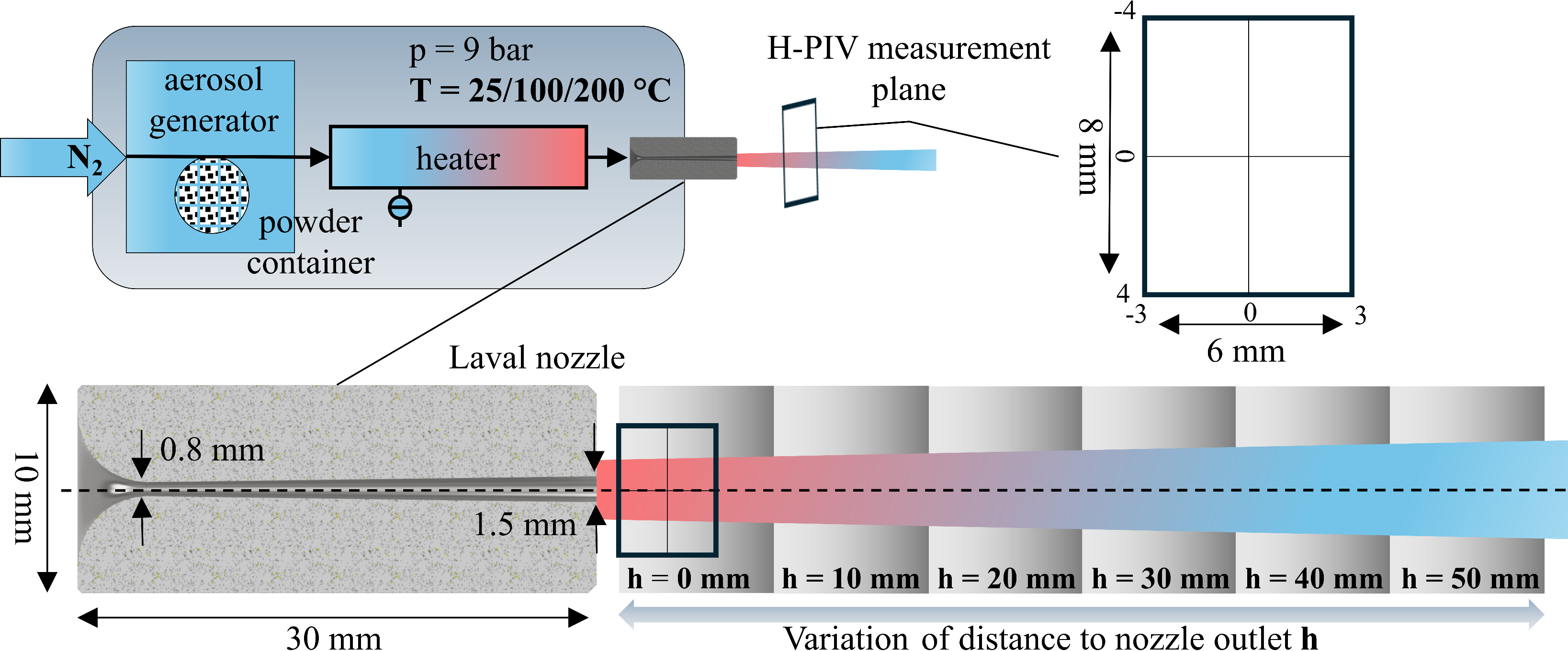}
    \caption{Illustration of the functional principle of the LPCGS setup and the measuring procedure during the parameter variation of the process gas temperature~$T$ and the distance of the measuring window from the nozzle outlet~$h$.}
    \label{CGS_SetUp}
\end{figure}

\begin{figure}  [htbp]
\centering
\begin{subfigure}{0.5\textwidth}
    \includegraphics[width=\textwidth]{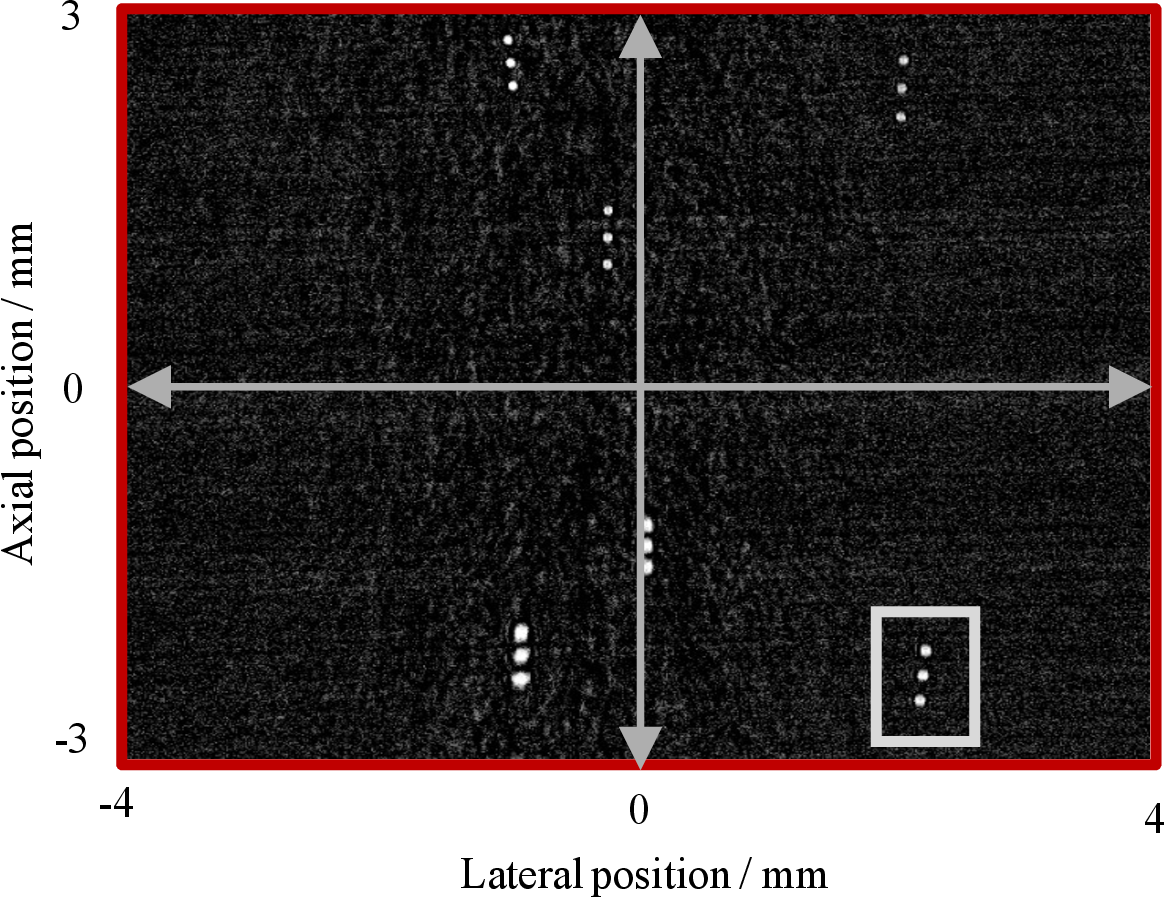}
    \caption{H-PIV image}
    \label{Triplet_Image}
\end{subfigure}
\hfill
\begin{subfigure}{0.3\textwidth}
    \includegraphics[width=\textwidth]{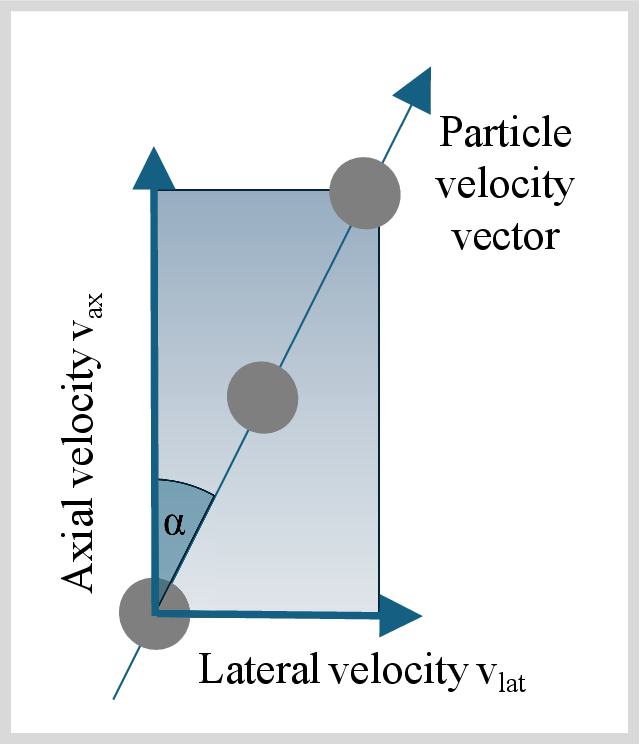}
    \caption{Trajectory angle $\alpha$}
    \label{Trajectory}
\end{subfigure}
\hfill
\caption{Representation of an H-PIV image with the characteristic shadowing triplets in a) and schematic representation of the particle trajectory angle in b).}
\label{subfigure_Hiwatch}
\end{figure}

The spray experiments are performed with the previously characterized feedstock powders on a LPCGS setup developed at the institute for the purpose of substrate surface microstructuring~\cite{Breuninger.2019b, 10.1007/978-3-031-44603-0_6, Thielen.2021, Zhu.2024, Buhl.2015}. The setup in Figure~\ref{CGS_SetUp} is operated with nitrogen as carrier gas at a pressure of 9~bar and a temperature range of $T$~=~25°C to 200°C. Aerosol generation is performed with a ring dosing powder feeder (Topas SAG 409, Topas GmbH, Germany) pressurized in a stainless steel tank. The dosing ring feeder operates with a powder container from which the feedstock powder is transferred to the rotating dosing ring via a screw conveyor. During rotation, the applied powder is transported to a venturi nozzle where it is dispersed with the carrier gas. Dosing via a rotating ring allows for uniform and pulse-free conveying of the powder, resulting in a constant particle load in the aerosol. The rotating speed of the dosing ring is kept constant for all powders and experiments. The resulting powder mass flows are shown in Table~\ref{tab_1_1}. 
In a subsequent step, the aerosol is heated by a tube heater before it is expanded over the Laval nozzle, where the particles are accelerated in the divergent area of the nozzle and in the free jet as long as the ambient gas velocity $v_{g}$ exceeds the particle velocity $v_{p}$ (cf. Figure~\ref{CGS_SetUp}). The axial exit velocity of the gas~$v_{g, exit}$ is mainly governed by the pressure difference between the inlet and outlet of the Laval nozzle, as well as the inlet temperature~$T_{in}$ (cf. Equation~(\ref{eq_laval})) \citep{King2015}.
\begin{equation}
\label{eq_laval}
    v_{g, exit}=\sqrt{\frac{T_{in}R}{M}\cdot \frac{2\kappa}{\kappa-1}\cdot\left( 1-\frac{p_{out}}{p_{in}}^{\frac{\kappa-1}{\kappa}} \right)}
\end{equation}
In addition to the differential velocity, the particle acceleration is determined by the drag coefficient~$c_w$, which is a function of the particle Reynolds number~$Re_p$ and the particle sphericity~$\Psi$ \citep{Haider.1989, Ganser.1993, Holzer.2008}. The nozzle geometry used is a special design with a throat diameter of 0.8~mm and an outlet cross section of 1.5~mm with a total length of 30~mm, which allows an optimal acceleration of particles in the size range of 1 - 10~µm~\citep{Buhl.2018}. The influence of particle shape is known to increase with the Reynolds number of the particles~\citep{Haider.1989}. Considering the correlation between the Reynolds number, size, and relative velocity of the particle, it can be inferred that particle shape has a greater impact on the movement of particles within the jet for larger particles that exhibit higher relative velocities with respect to the gas. Previous simulations of this Laval nozzle~\citep{Breuninger.2022} have shown that spherical steel particles within the size range of 1 to 25~µm achieve maximum particle Reynolds numbers in the throat of the nozzle used, spanning from 20 to 980. These simulations further demonstrated that the change in the shape of the particles with a size range of 10 to 25~µm exerted the most significant influence on the particle velocity~\citep{Breuninger.2022, Antonyuk.2024}.
\begin{table}[htbp]
\centering
\caption{Powder mass flows $\dot{m}$ of the feedstock powders at constant dosing ring speed $\omega$.}
\begin{tabular}{lcll}
                                   & CuCR         & \multicolumn{1}{c}{CuSA} & \multicolumn{1}{c}{CuTSF} \\ \hline
Ring speed \\ $\omega$ / rpm            & \multicolumn{3}{c}{1.64 $\pm$ 0.07}                                    \\
Mass flow \\ $\dot{m}$ / mg s\textsuperscript{-1} & 3.88 $\pm$ 0.39 & 2.85 $\pm$ 0.33             & 2.51 $\pm$ 0.14               \\ \hline
                                   &              &                          &                          
\end{tabular}
\label{tab_1_1}
\end{table}
During the spray experiments, an H-PIV system (HiWatch HR2, Oseir Ltd., Finland) is used to determine the particle velocity and spatial distribution in the spray jet. The measurement principle is based on the obscuration of laser pulses by passing particles in the measurement field (8x6~mm$^2$ with 400~µm field depth). The measuring window of the H-PIV system is positioned parallel to the direction of flow of the nozzle stream. The measuring system is aligned so that the spray cone is aligned with the zero line of the measuring window in the lateral position (cf. Fig.~\ref{Triplet_Image}). When the back-illuminated image sensor (Complementary Metal Oxide Semiconductor (CMOS)) is exposed individually, the shadowing of the laser pulses leads to the formation of shadow triples in the image (cf. Fig.~\ref{Triplet_Image}). Based on the selected pulse duration and interval length, the particle velocity and position can be calculated from the generated image data. Furthermore, the triplets allow the characterization of the particle diameter in the size range of 5-1000~µm. By adjusting the laser interval and pulse duration down to the nanosecond range, particle velocities up to 2000~m/s can be tracked~\cite{Koivuluoto.2018}. The measurements were taken with a pulse duration of 80~ns and a pulse interval length of 800~ns. As described in Figure~\ref{CGS_SetUp}, the H-PIV measurements are performed at varying distances from the nozzle exit $h$ in 10~mm intervals up to a maximum distance of 50~mm. The indication of the distance to the nozzle $h$, refers to positioning of the 8~mm edge of the 8x6~mm measuring field closest to the nozzle, as illustrated in Figure~\ref{CGS_SetUp}. When specifying the nozzle distance $h$, the analysis of particle velocities in a range from $h + 6$~mm is therefore considered. For accurate positioning, the HiWatch HR2 system is placed on a 3-axis positioning stage.
The particle velocity is output in both magnitude and components of the vector, namely the axial velocity component $v_{ax}$ and the lateral velocity component $v_{lat}$ (cf. Fig.~\ref{Trajectory}). This allows the calculation of the particle trajectory angle $\alpha$ with respect to the zero line of the measurement range in the lateral position. To complete the parameter study, the measurements are supplemented by a carrier gas temperature variation from 25°C over 100°C and 200°C. 


\section{Results and Discussion}
\label{sec3}

\subsection{Analysis of the H-PIV measurements}
\label{subsec3_1}

\begin{figure}  [htbp]
\centering
\begin{subfigure}{0.65\textwidth}
    \includegraphics[width=\textwidth]{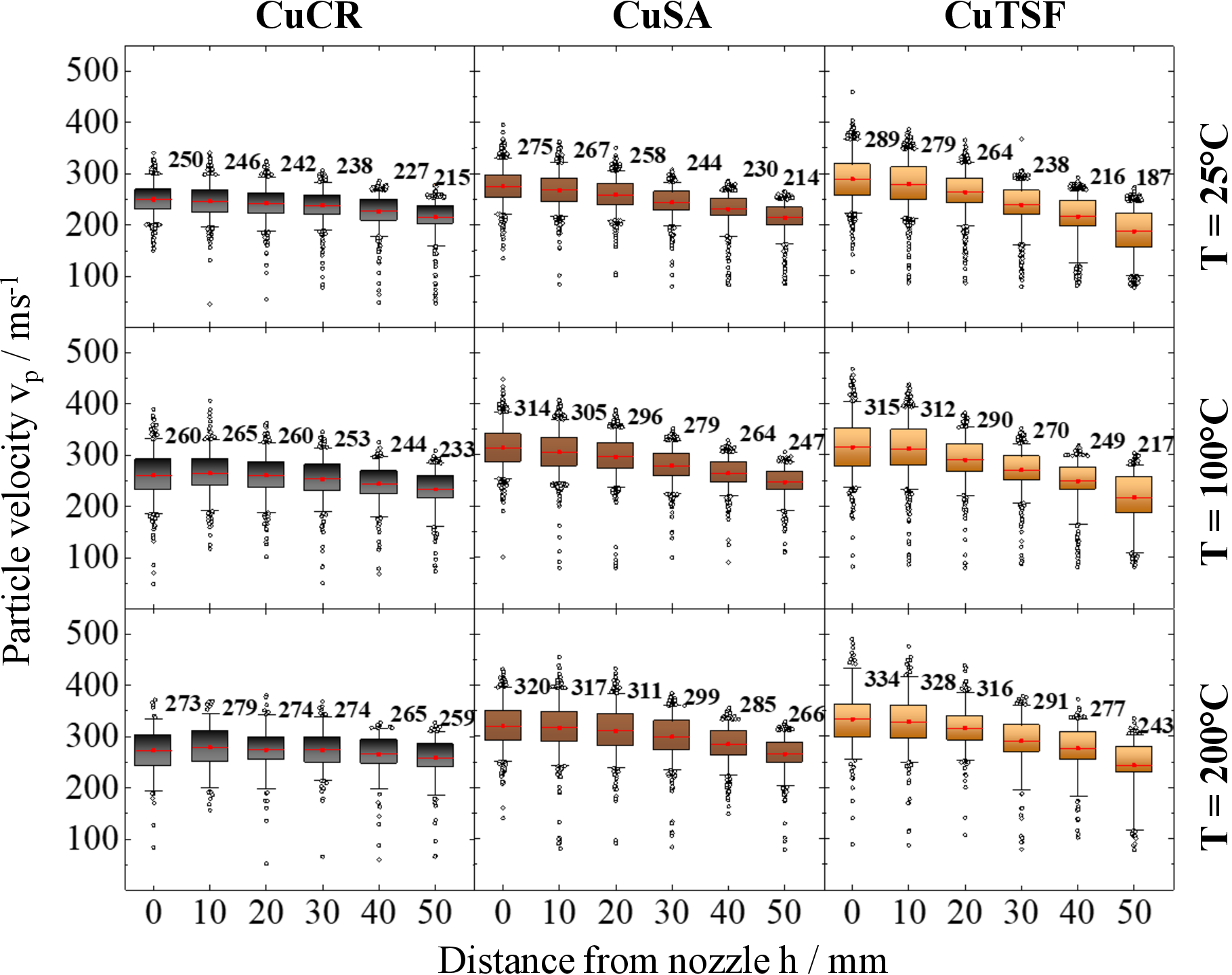}
    \caption{Particle velocity distributions}
    \label{Particle_Velocity}
\end{subfigure}
\\[2ex] 
\begin{subfigure}{0.65\textwidth}
    \includegraphics[width=\textwidth]{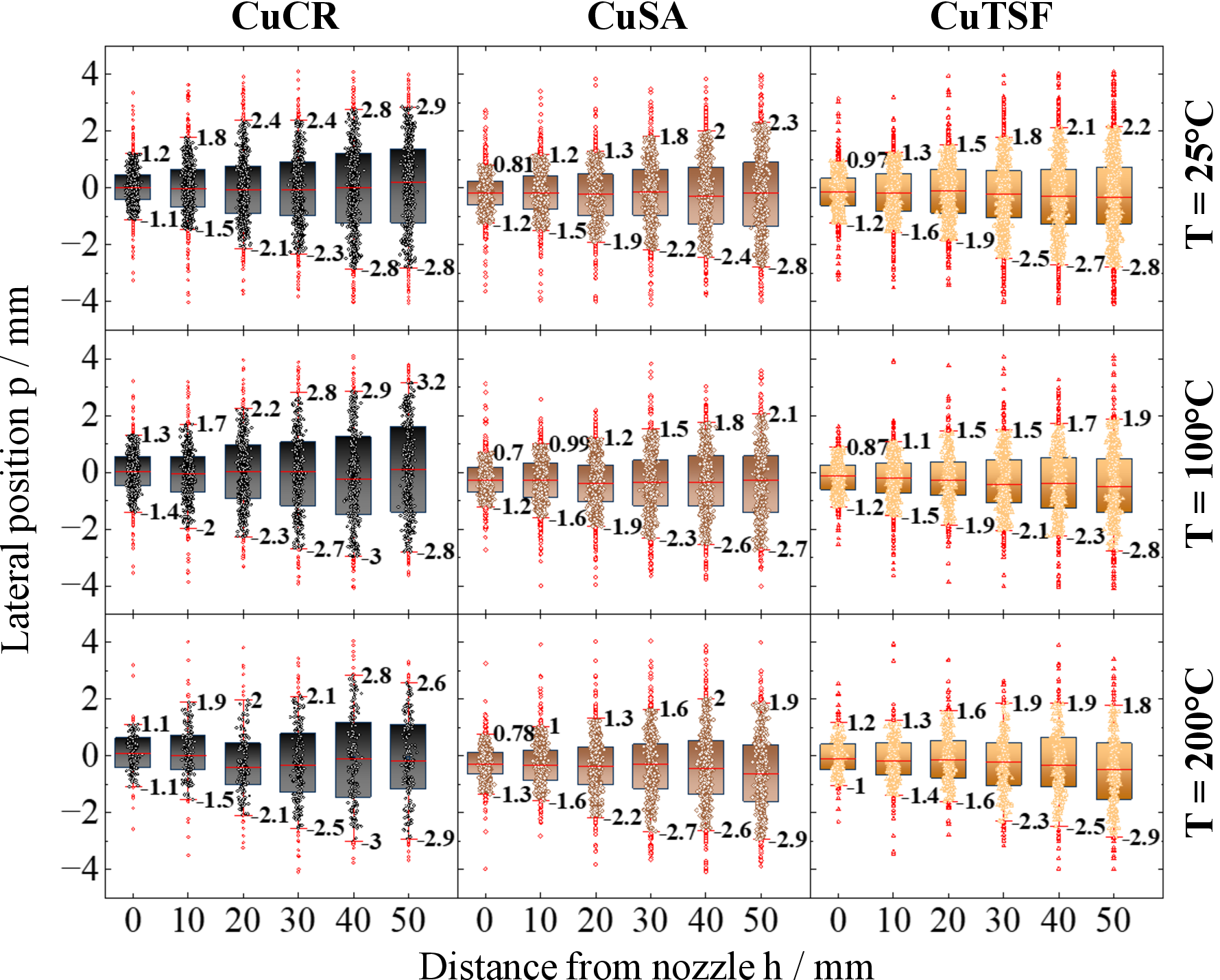}
    \caption{Particle location distributions}
    \label{Particle_Position}
\end{subfigure}
        
\caption{Representation of the velocity and spatial distributions of the particles for the investigated parameter space, displayed as a box plot. The median value of the distribution per parameter set is shown as a red line in the box plots. The top of the box indicates the 75th percentile of the distribution and the bottom of the box indicates the 25th percentile of the distribution. The whiskers indicate the 95th and 5th percentile of the distribution.}
\label{Subfigures_Vel_Pos}
\end{figure}

\begin{figure} [htbp]
    \centering
    \includegraphics[width=\linewidth]{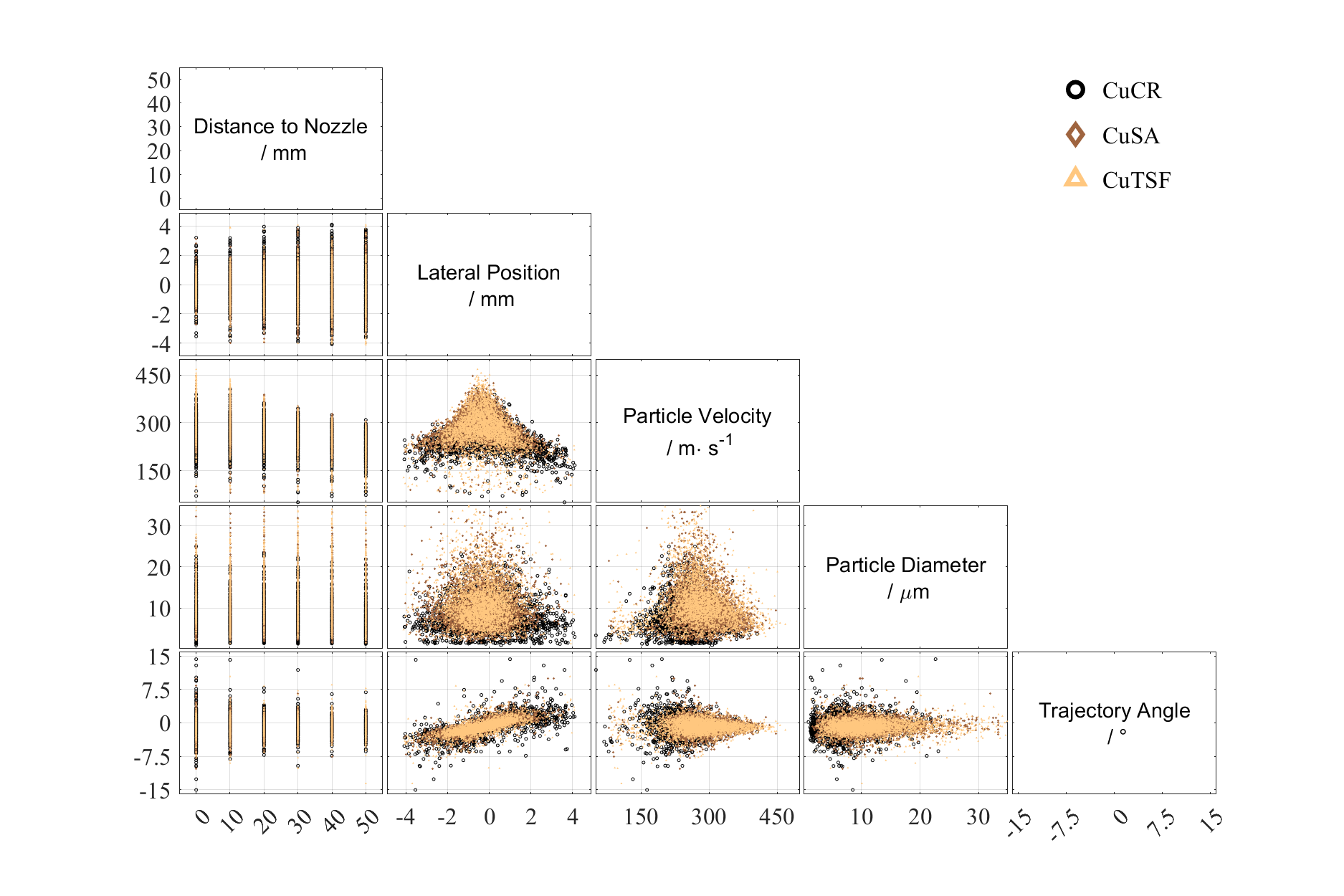}
    \caption{Display of H-PIV measurement and analysis data as a plot matrix at a gas temperature of $T$~=~100°C. The corresponding measured quantities are shown on the diagonal. The respective scatter plots show the entirety of the measurement data under variation of the particle material (marker color) and the distance to the nozzle $h$.}
    \label{fig_3_1}
\end{figure}

\begin{figure}  [htbp]
\centering
\begin{subfigure}{0.49\textwidth}
    \includegraphics[width=\textwidth]{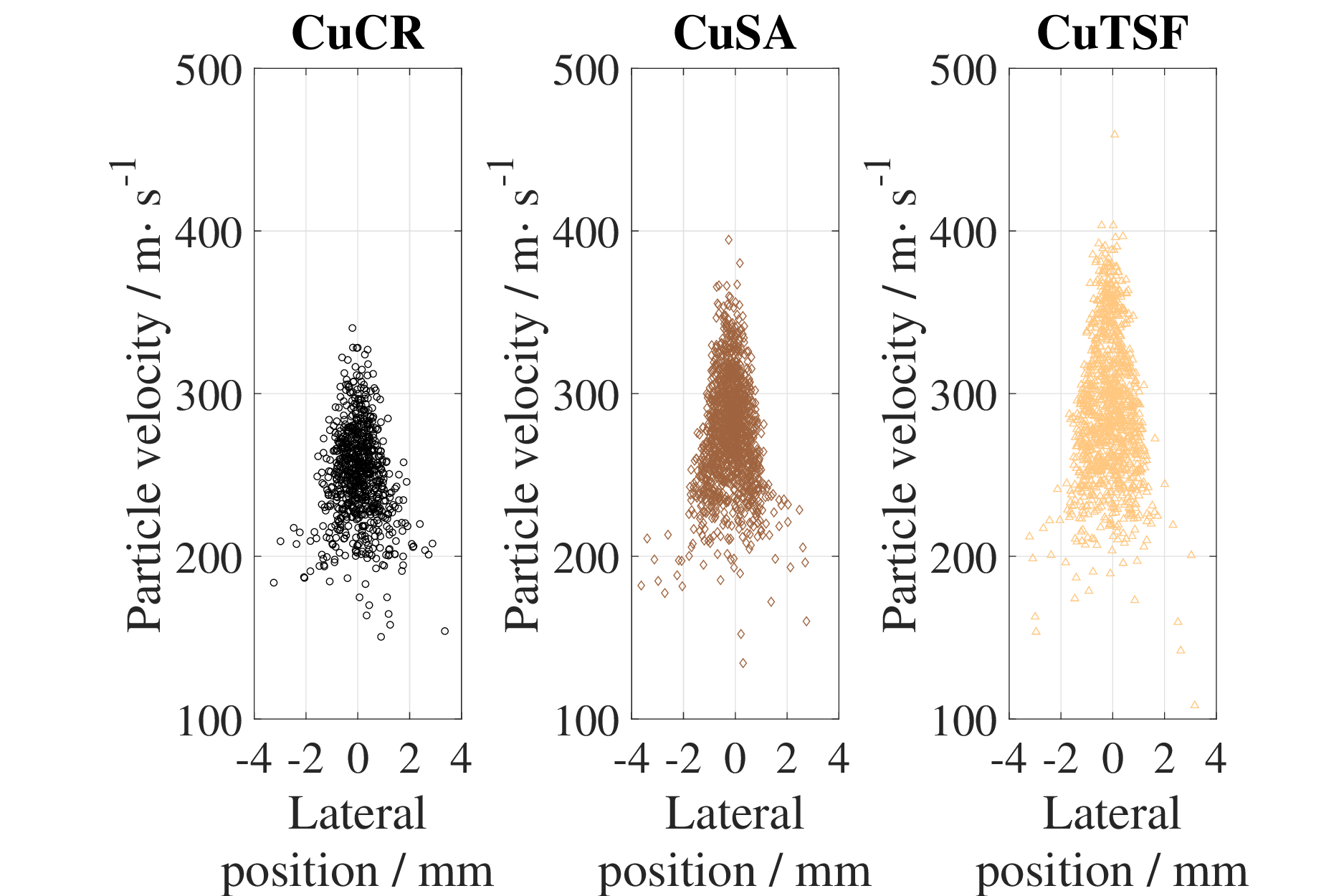}
    \caption{Particle velocity over lat. position}
    \label{Velocity_Position}
\end{subfigure}
\vspace{0.25em}
\begin{subfigure}{0.49\textwidth}
    \includegraphics[width=\textwidth]{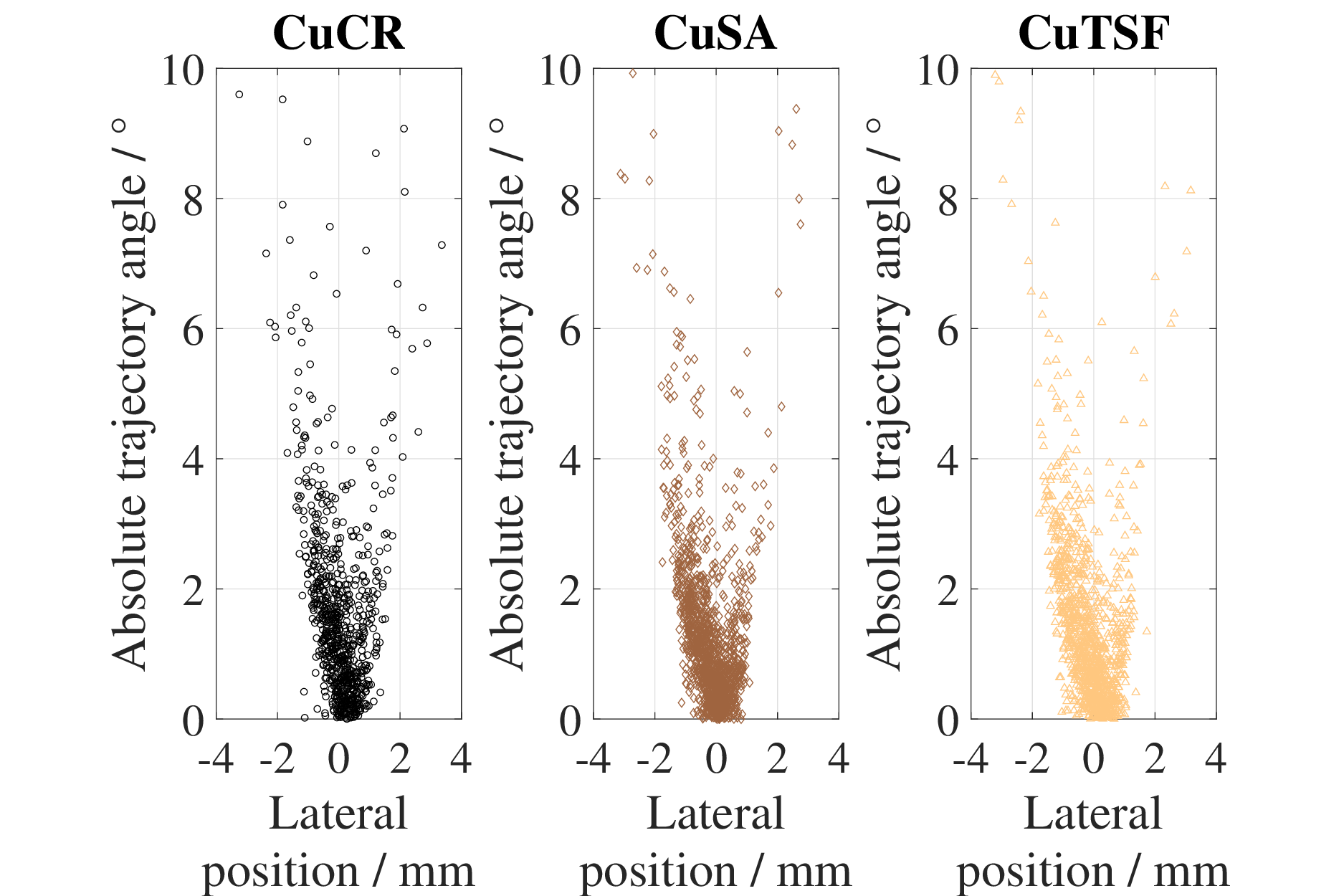}
    \caption{Abs. traj. angle over lat. position}
    \label{Angle_Position}
\end{subfigure}
\hspace{1em}
\begin{subfigure}{0.49\textwidth}
    \includegraphics[width=\textwidth]{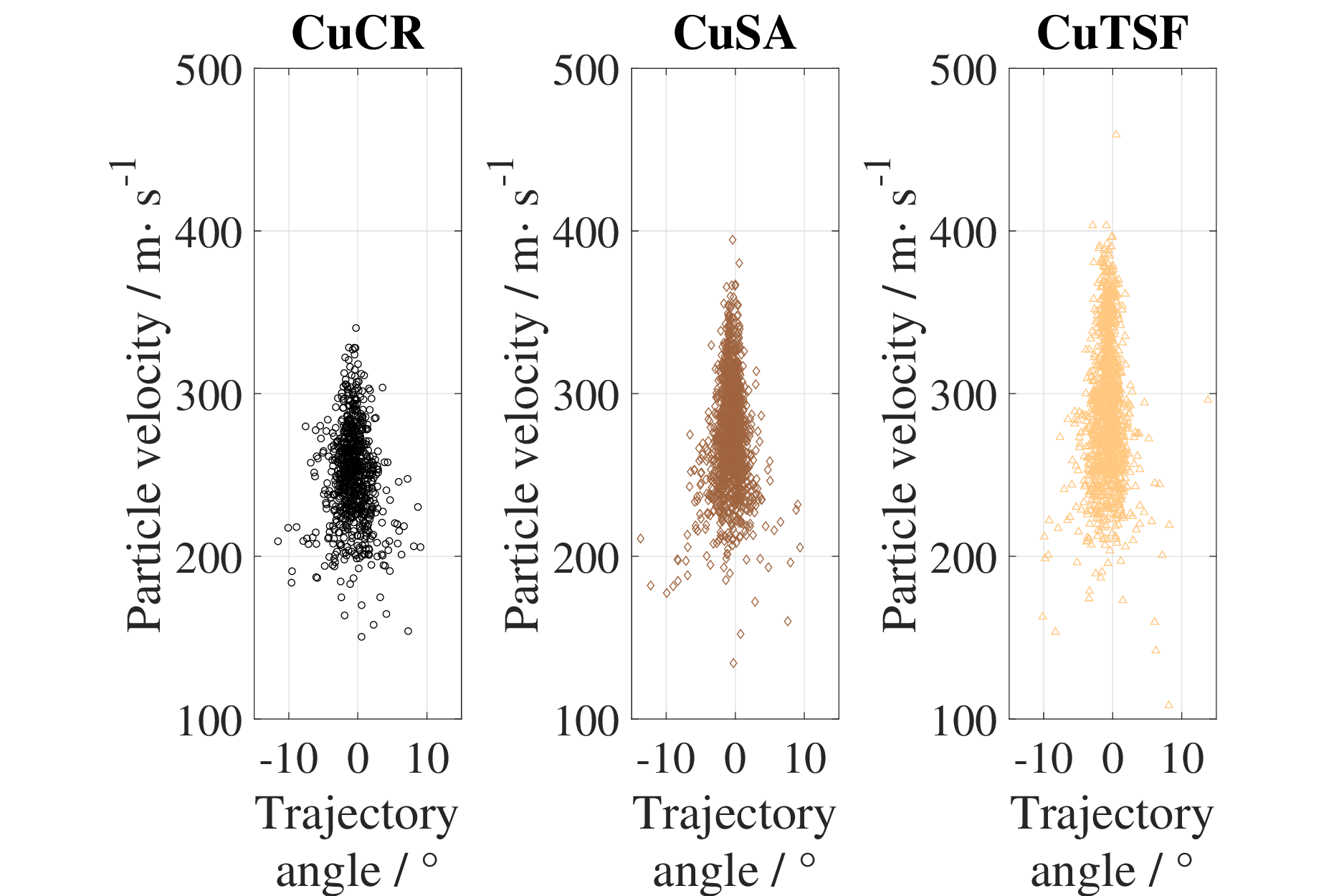}
    \caption{Velocity over trajectory angle}
    \label{Velocity_Angle}
\end{subfigure}
\vspace{0.25em}
\begin{subfigure}{0.49\textwidth}
    \includegraphics[width=\textwidth]{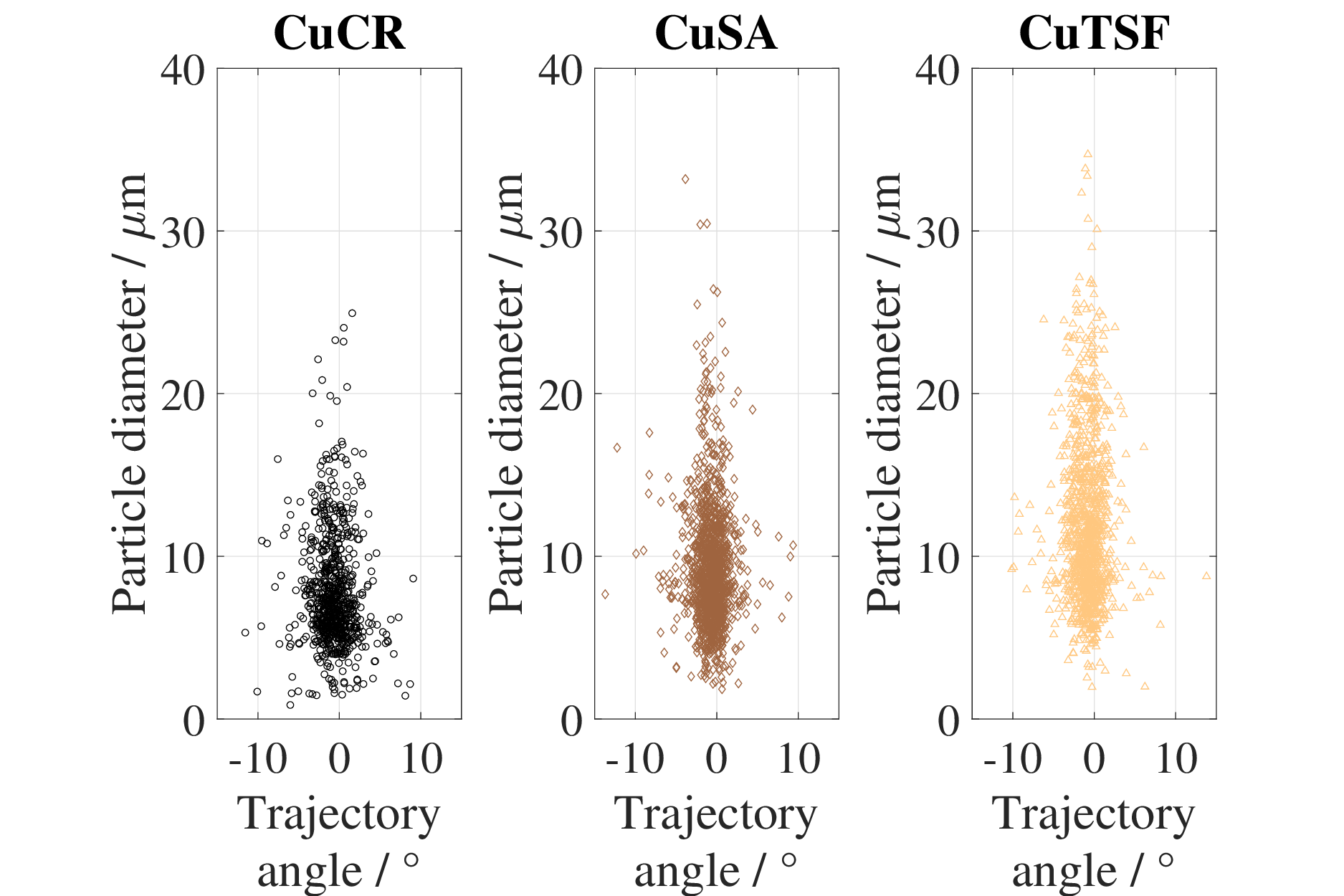}
    \caption{Particle diameter over trajectory angle}
    \label{Particle_Angle}
\end{subfigure}
        
\caption{Relationship between the particle velocity and the position in the spray jet (a), as well as the trajectory angle (c), at the nozzle outlet $h$~=~0 at a process temperature of $T$~=~25°C. In (b), the absolute trajectory angle is plotted against the lateral position and in (d) the particle diameter is plotted against the particle trajectory angle.}
\label{subfigure_anglevslatpos}
\end{figure}

The analysis of the H-PIV measurements enables the determination of the velocity, size, and position of the particles, as well as their correlation with the previously described morphological properties. The results of the parameter study for the investigated particle systems are shown in Figure~\ref{fig_3_1} as an example for a process gas temperature of $T$~=~100°C. The determined data points are shown as a plot matrix, with the respective measured quantities on the diagonal. For completeness, the plot-matrices of parameter variation at $T$~=~25°C and $T$~=~200°C are shown in the Appendix~ Fig.~\ref{multivariant_allT25_app} and Fig.~\ref{multivariant_allT200_app}.\\ 
Given that the particle impact velocity is a critical factor determining the deposition efficiency and deformation of particles in the cold gas process, the quantification of the particle velocities in the spray cone is of particular interest. The particle velocity distributions and spatial distributions measured during the parameter variation are presented in Figure~\ref{Subfigures_Vel_Pos} as box plots, with the data illustrated over the distance to the nozzle outlet~$h$ for each case. The outcomes for the copper powders utilized are presented in a column-by-column manner, with increasing process gas temperatures from top to bottom. The median of the velocity distribution (cf. Figure~\ref{Particle_Velocity}) demonstrates a dependence on the nozzle distance~$h$ for all particle systems. The highest velocities are observed for irregular particle systems (CuTSF, CuSA) at low distances $h$. This phenomenon can be attributed to the influence of the drag coefficient, which is higher for irregularly shaped particles and thus leads to a higher acceleration of the particles~\citep{Michaelides.2023}.
Conversely, this phenomenon also pertains to the deceleration of the particles if the velocity of the surrounding gas is lower than that of the particles. Furthermore, the particle velocity data in Figure~\ref{Particle_Velocity} are showing a wider distribution with decreasing sphericity. This can especially be seen, when comparing the velocity distributions of CuCR and CuTSF. In addition, the curve of the particle velocity over the nozzle distance shows a less constant course for particle systems with low sphericity. As in the case of higher acceleration in the nozzle, particles with an irregular shape are decelerated more quickly when the velocity of the surrounding fluid decreases with increasing distance from the nozzle exit (cf. Figure~\ref{Particle_Velocity}). This phenomenon elucidates the pronounced decline in median particle velocity observed for CuSA and CuTSF. Conversely, the spherical CuCR particles exhibit a more constrained velocity distribution, demonstrating reduced reliance on the distance to the nozzle~$h$. Nevertheless, the maximum attainable particle velocities are diminished, and the distribution of particle velocities is more confined in general. The findings from the analysis of the measurement data at 100°C are transferable to the measurements at T~=~25°C and T~=~200°C. As the process gas temperature is increased, the overall particle velocity also increases, due to the increase in gas velocity~\citep{Breuninger.2022}. This influence can be observed more strongly for irregularly shaped particles, which can be explained by the growing influence of the shape with increasing velocity or Re-number~\citep{Haider.1989}. Nevertheless, the rise in temperature does not result in substantial changes in the spatial distribution of the particles in the jet (cf. Fig.~\ref{fig_3_1}, Fig.~\ref{multivariant_allT25_app} and Fig.~\ref{multivariant_allT200_app}).
 As demonstrated in Figure~\ref{fig_3_1}, a clear distinction can already be made between the examined particle systems with regard to the variables investigated. Specifically, differences in particle velocity and focusing, defined as the spatial disposition of the particles, can be determined.\\ 
In addition to the velocity distribution of the particles, the spatial distribution of the particles within the spray is also crucial for optimizing the production process, since it determines the final structure resolution. Consequently, it is imperative to seek the highest attainable resolution by focusing the particle stream, as this is instrumental for enhancing the efficiency of the production process. The degree of focusing is contingent upon the geometric parameters of the Laval nozzle, particularly the outlet diameter~\citep{Sova.2013, Sova.2013b, Sova.2018}. To achieve the smallest possible jet width and thus focus the particles, the throat diameter must be adjusted accordingly. However, this adjustment imposes constraints on the manufacturing process due to the inherent limitations of the process. The nozzle utilized in the experimental phase, with an outlet diameter of 1.5~mm and a throat diameter of 0.8~mm, is positioned at the lower end of the nozzle size range.
In this regard, the analysis of the particle position within the spray cone demonstrates a notable correlation. When plotting the particle positions against the distance to the nozzle~$h$ in Figure~\ref{Particle_Position}, it becomes observable that the distribution width is found to be strongly dependent on the particle morphology. Taking the particle position distribution at $T$~=~100°C and at a distance of $h$~=~0 as an example, the difference in the measured spray width between the spherical CuCR and the aspherical CuSA and CuTSF powders is 30\%.
Consequently, an increasing focusing of the particle beam can be observed for particles exhibiting higher irregularity. This finding presents the potential prospect of manipulating the spray resolution in a targeted manner through the systematic selection of the particle system. Meanwhile, these findings can be established for all nozzle spacings and particle systems at the investigated process gas temperatures. Even at low nozzle distances~$h$ of 0 to 20~mm, there are strong differences in the focusing of the particles in the copper particle systems investigated. As this range corresponds to the working range of cold gas spraying, it can be assumed that the particle morphology has an influence on the focusing in the spray beam, which should be transferred to the structure resolution in typical cold gas spraying applications. \\ 
Furthermore, the H-PIV data can be used to determine the particle trajectories based on the components of the velocity vector as shown in Figure~\ref{Trajectory}. The direction of movement of the particles can thus be determined in relation to the deviation of the zero line in lateral position. The particle trajectory vector can be utilized to ascertain the angle to the zero line of the measuring range, employing trigonometric relationships. As the zero line corresponds to the axis of symmetry of the Laval nozzle, a statement can be made about the focusing of the particles in the spray jet, which in turn determines the production resolution of the process. Figure~\ref{subfigure_anglevslatpos} illustrates the relationships between the velocity of particles and their position or trajectory angle. It is evident from the graph that particles reach their maximum velocities at the center of the jet, while the lowest velocities are observed in the edge areas (cf. Figure~\ref{Velocity_Position} and \ref{Velocity_Angle}). The data presented here were evaluated in a parameter study with a distance to the nozzle of $h$~=~0, thereby representing the behavior of particles immediately after exiting the nozzle. The decrease in particle velocities at the edge of the spray cone can be attributed, in part, to the flow conditions in the nozzle, where the gas velocity decreases towards the wall \citep{Buhl.2018, Breuninger.2019, Breuninger.2019b}. Furthermore, the turbulence increases from the center of the jet to its outer hull, which also influences the particle trajectories \citep{Yu.2023}.
Moreover, the illustration enables the analysis of the trajectory angle dependence from the particle position (Fig.~\ref{Angle_Position}). As the distance to the symmetry axis of the spray cone, delineated by the zero line in lateral position, increases, the particles' trajectory angle exhibits a marked increase relative to particles that are centered within the spray jet. The data points demonstrate a uniform increase in the angles towards the edge; no area with a sudden increase is discernible. Moreover, particles exhibiting a lower trajectory angle demonstrate higher velocities~(Fig.~\ref{Velocity_Angle}). The measurements also show that these particles, located at the centre of the jet, have larger diameters~(Fig.~\ref{Particle_Angle}). Conversely, particles of smaller size demonstrate faster response to changes in gas flow conditions due to their comparatively lower inertia and exhibit the highest trajectory angles. These relationships are consistent with the outcomes previously obtained from the simulations of this nozzle~\citep{Aleksieieva.2024}.
\begin{figure}  [htbp]
\centering
\begin{subfigure}{0.45\textwidth}
    \centering
    \includegraphics[width=\textwidth]{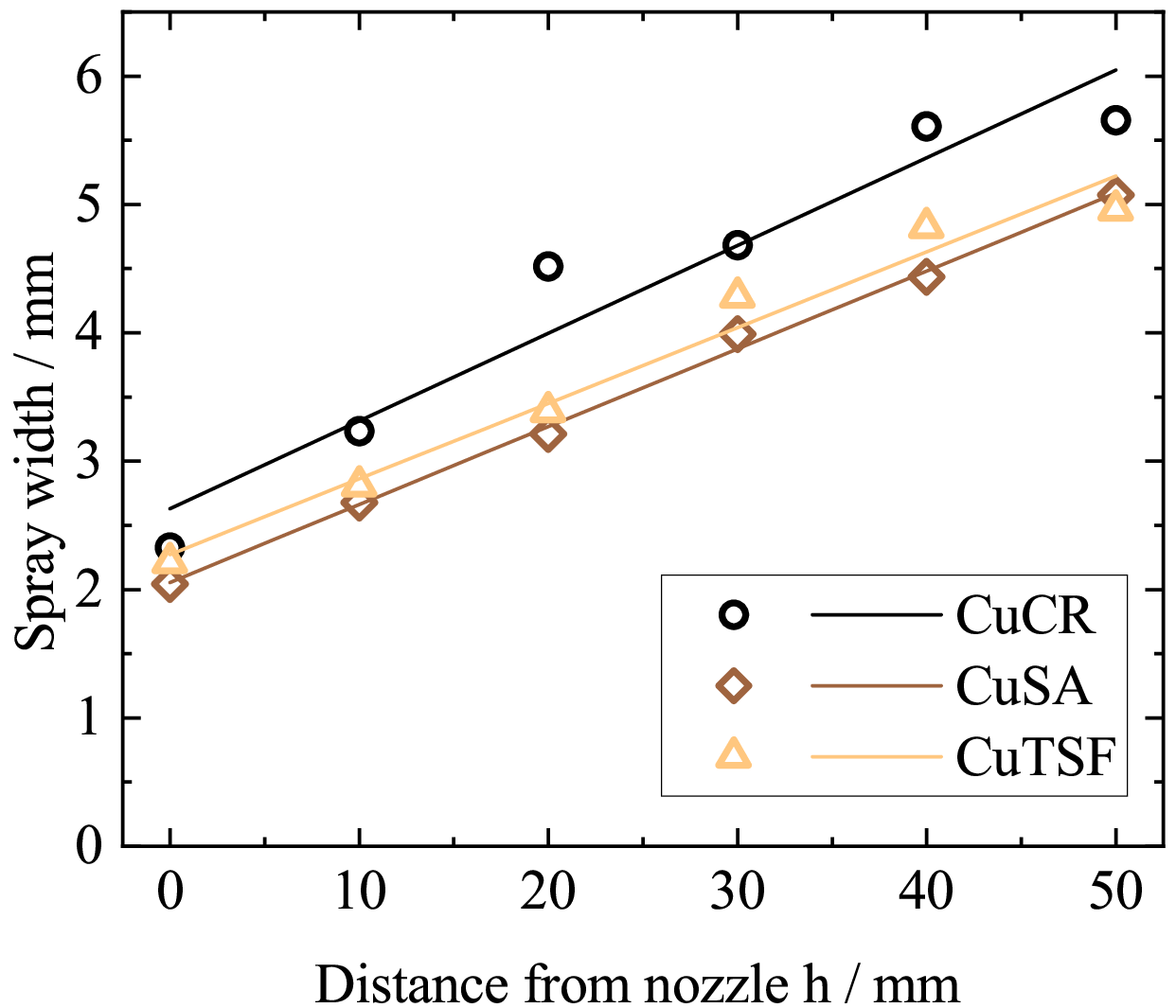}
    \caption{Spray cone width at T = 100°C}
    \label{spraywidth}
\end{subfigure}
\hfill
\begin{subfigure}{0.45\textwidth}
    \centering
    \includegraphics[width=\textwidth]{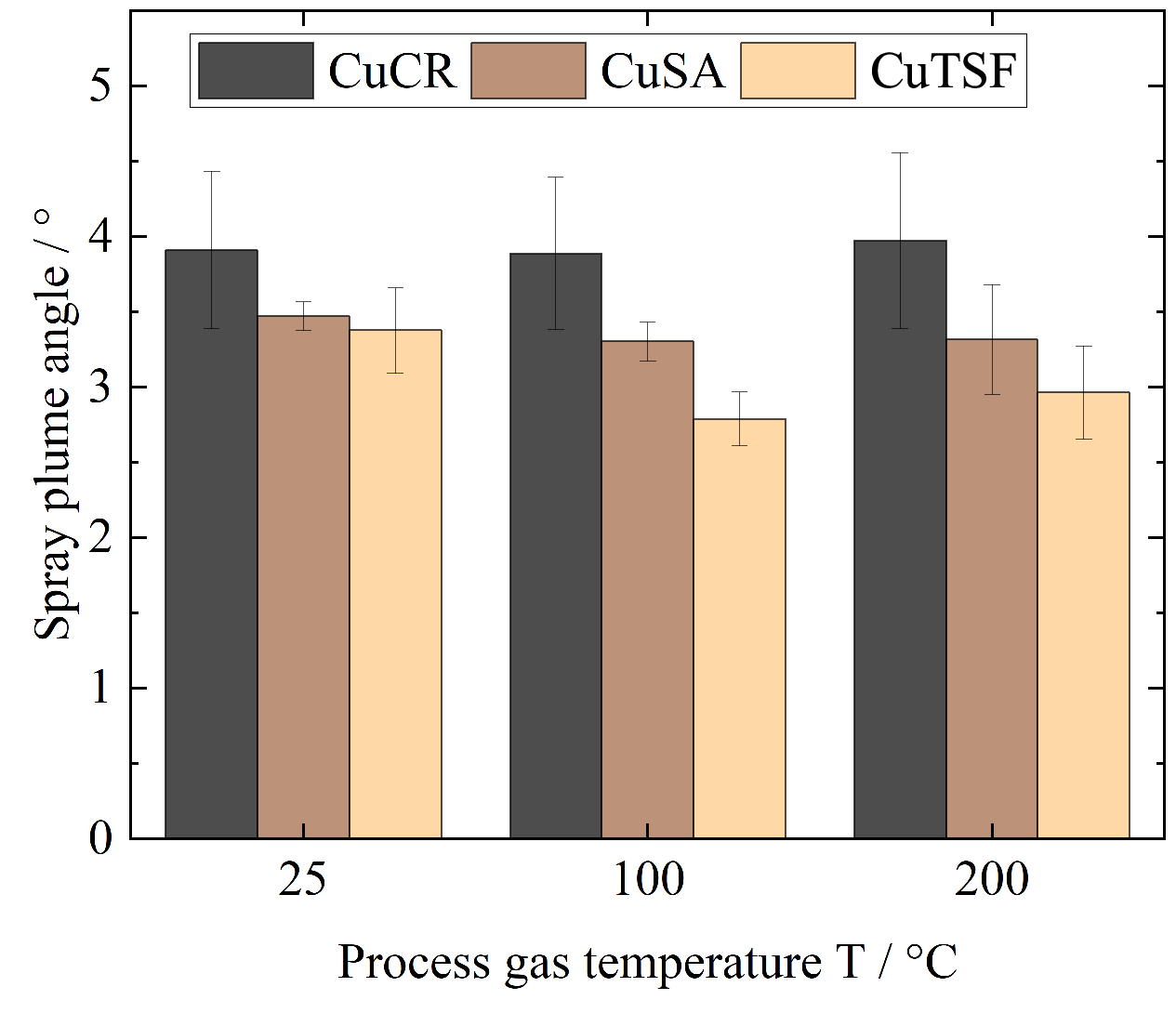}
    \caption{Spray cone angle}
    \label{sprayangle}
\end{subfigure} 

\caption{Representation of the spray width in dependence of the nozzle distance $h$ at $T$~=~100°C in (a) and the resulting spray cone angels in (b) for the investigated process the investigated process gas temperatures at the nozzle inlet.}
\label{spray_angle}
\end{figure}
The findings underline the significant effect on the distribution of the particles in the spray jet, as demonstrated in Figure~\ref{spray_angle}. When the spray widths, as measured from the spatial particle distribution~(cf.~Fig.\ref{Particle_Position}), are analyzed and plotted against the distance to the nozzle, a linear relationship becomes evident. Based on this relationship, the spray cone angles for the respective particle system are calculated. A further examination of the dependency of the spray angles for each particle collective reveals a clear dependency on the morphology~(Fig.~\ref{sprayangle}). However, the analysis did not reveal any discernible temperature dependence for the examined particle systems within the considered temperature range.

\section{Conclusion}
\label{sec4}

In the present study, the impact of the morphology of copper microparticles on their spray characteristics during the low-pressure cold gas process was investigated. Employing 2D and 3D imaging methods enabled the precise quantification of the particle's shape and size. Static image analysis facilitates the measurement of the shape and size of particle systems with minimal effort. The validity of this method was assessed by X-ray tomographic imaging of the particle collectives. The technique allows for the accurate determination of the surface area and volume of the particles. A comparison of the 3D X-ray with the 2D image analysis demonstrates that the 2D method provides a reliable indication of the particle shape and size. However, 3D reconstruction of particle collectives provides extended structural data, especially for particle systems with porous structure and concave surface topography, which cannot be sufficiently quantified by 2D methods (cf.~ CuTSF~in~Fig.\ref{fig_2_1a}~and~Fig.\ref{fig_2_3}). \\ 
In order to analyze the influence of the particle morphology on the spray behavior, H-PIV measurements were carried out in the free jet. For this purpose, the process gas temperature and the measuring position were varied based on the outlet position of the Laval nozzle. The H-PIV measurements allow for the quantification of the particle velocity as well as their movement direction and size. The increase in temperature results in an increase in particle velocity, particularly for irregularly shaped particles, while the spatial distribution of particles within the jet remains unaffected. As the distance from the nozzle increases, particle velocity decreases, with this effect becoming notably more pronounced as the particle shape deviates further from a sphere. The radial distribution of particles within the spray cone is governed by particle size: smaller particles tend to be deflected toward the edges of the cone, whereas larger particles are concentrated near the center of the jet due to their inertia. However, the distribution of particles by diameter becomes significantly narrower as their sphericity decreases. The analysis of the obtained data revealed a clear correlation between the increase in particle irregularity and the corresponding increase in velocity and focusing within the jet. The results obtained in this study can be used for the further optimization of the low-pressure cold gas process in a range of applications, including repair, coating and additive manufacturing~(AM) methods. 

\section{Acknowledgement}
\label{acknowledgement}

The authors gratefully acknowledge funding by the German Research Foundation (DFG) within SPP 2364 under
grant 504954383.




\clearpage

\appendix
\section{Appendix}
\label{app1}

\begin{figure} [htbp]
    \centering
    \includegraphics[width=\linewidth]{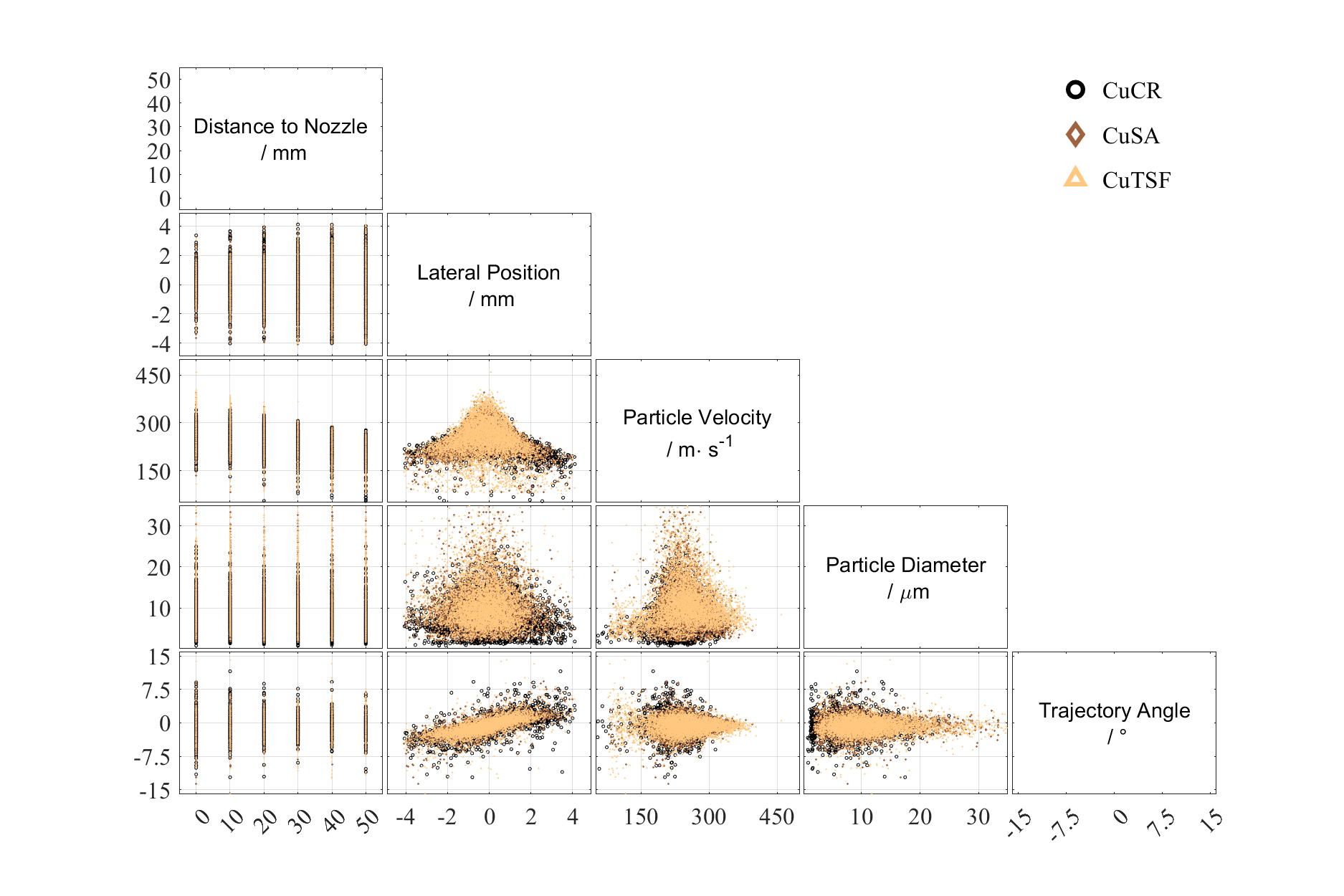}
    \caption{Display of H-PIV measurement and analysis data as a plot matrix at a gas temperature of $T$~=~25°C. The corresponding measured quantities are shown on the diagonal. The respective scatter plots show the entirety of the measurement data under variation of the particle material (marker color) and the distance to the nozzle $h$.}
    \label{multivariant_allT25_app}
\end{figure}

\begin{figure} [htbp]
    \centering
    \includegraphics[width=\linewidth]{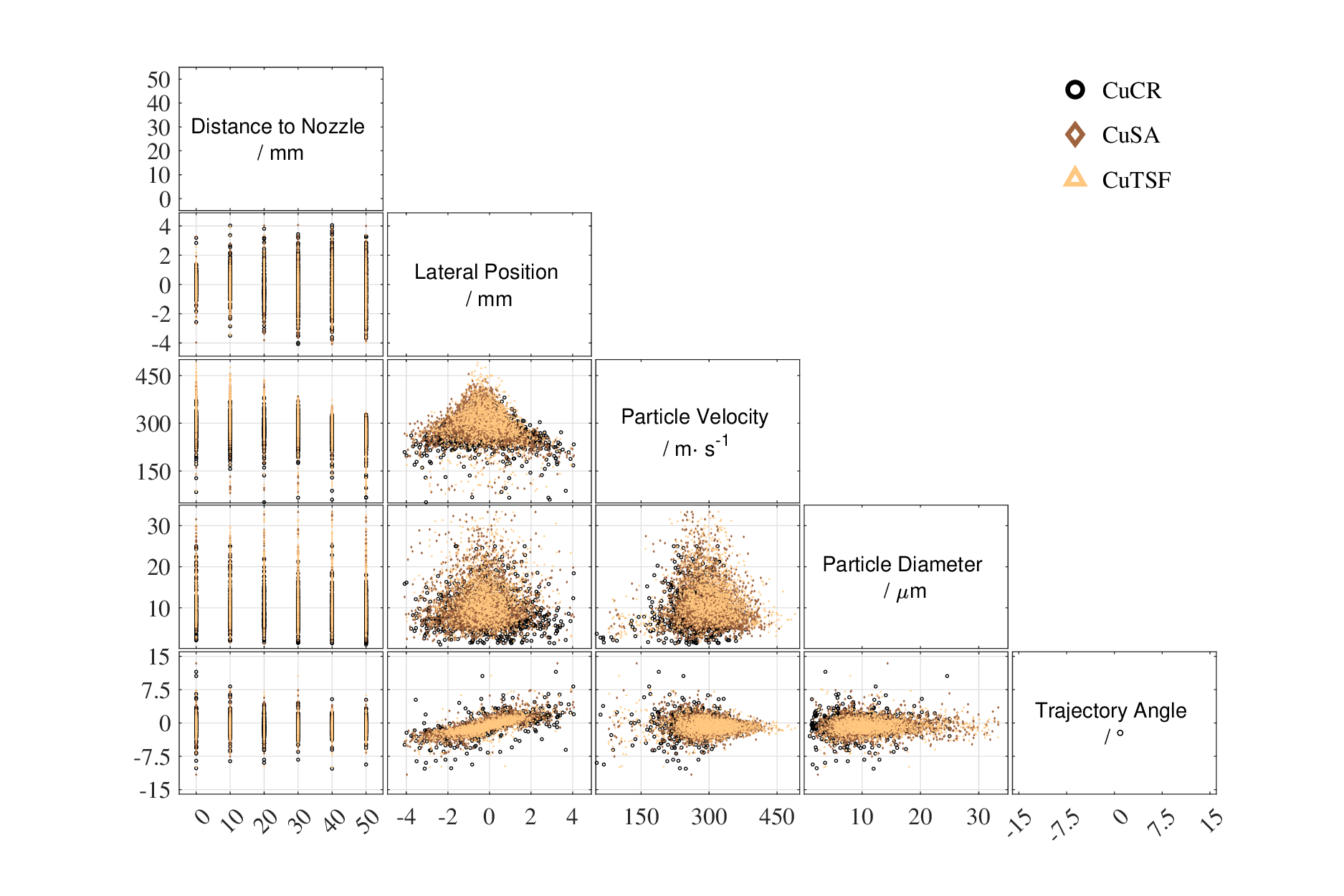}
    \caption{Display of H-PIV measurement and analysis data as a plot matrix at a gas temperature of $T$~=~200°C. The corresponding measured quantities are shown on the diagonal. The respective scatter plots show the entirety of the measurement data under variation of the particle material (marker color) and the distance to the nozzle $h$.}
    \label{multivariant_allT200_app}
\end{figure}



\newpage

\bibliographystyle{abbrvnat}  
\bibliography{Lit_MorphPaper}

\end{document}